\documentclass[AMA,STIX1COL]{WileyNJD-v2}

\articletype{Article Type}%

\usepackage{physics}
\usepackage{amsmath}
\usepackage{esint}
\usepackage{multicol}
\usepackage{pdflscape}
\usepackage{multirow}
\usepackage{booktabs}
\usepackage{subcaption}

\begin{document}

\title{Particle-resolved simulations of four-way coupled polydispersed, particle-laden flows}

\author[1,2]{Yinuo Yao*}
\author[3]{Edward Biegert}
\author[3,4]{Bernhard Vowinckel}
\author[3,6]{Thomas K\"{o}llner} 
\author[3]{Eckart Meiburg}
\author[5]{S. Balachandar}
\author[2]{Craig S. Criddle}
\author[1]{Oliver B. Fringer}

\address[1]{The Bob and Norma Street Environmental Fluid Mechanics Laboratory, Department of Civil and Environmental Engineering, Stanford University, Stanford, CA, 94305, USA}
\address[2]{Codiga Resource Recovery Center at Stanford, Department of Civil and Environmental Engineering, Stanford University, Stanford, CA, 94305, USA}
\address[3]{Mechanical Engineering, University of California, Santa Barbara, Santa Barbara, CA 93106, USA}
\address[4]{Leichtwei{\ss}-Institute for Hydraulic Engineering and Water Resources, Technische Universit\"{a}t, Braunschweig, 38106 Braunschweig, Germany}
\address[5]{Department of Mechanical and Aerospace Engineering, University of Florida, Gainesville, Florida 32611, USA}    
\address[6]{CADFEM GmbH, 85567 Grafing b. M\"{u}nchen, Germany}  

\abstract{We present a collocated-grid framework for Direct Numerical Simulations of polydisperse particles submerged in a viscous fluid. The fluid-particle forces are coupled with the Immersed Boundary Method (IBM) while the particle-particle forces are modeled with a combination of contact and lubrication models, adapted for collocated grids. Our method is modified from the staggered-grid IBM of previous authors to a collocated-grid IBM by adapting the fluid and particle solvers. The method scales well on high-performance parallel computing platforms. It has been validated against various cases and is able to reproduce experimental results. Tuning parameters have been thoroughly calibrated to ensure the accuracy of the method. Finally, we demonstrate the capability of the method to simulate both monodispersed and bidispersed fluidized beds and reproduce the power law relationship between the inflow velocity and the porosity.}


\corres{*Corresponding author name, \email{yaoyinuo@stanford.edu}}

\presentaddress{Department of Energy Resources Engineering, Stanford University, Stanford, CA, 94305, USA}
\maketitle

\section{Introduction}
Particle-laden flows are common in both industrial and natural systems. Typical examples of industrial applications in treating wastewater include sedimentation tanks and fluidized beds etc.~\cite{Shin2014-ad, Yao2021-ex, Yao2021-ky}, while examples of natural systems include sediment transport in estuaries~\cite{Fringer2019-gh}. Understanding the fluid-particle and particle-particle interactions enhances the design of various industrial systems. For instance, predicting the upflow velocity to achieve the desired porosity is critical in fluidized-bed applications. Hence, Richardson and Zaki~\cite{Richardson1954-ay} proposed a power law relationship to correlate the effect of superficial velocity (average inlet velocity) to the porosity of the particle phase by studying the macroscopic properties and behavior of the systems. However, due to limitations in the experimental data collection technologies, understanding and quantifying the microscale behavior has been a persistent challenge~\cite{Yin2007-eb}. In recent years, with improvements in both computational power and numerical methods, Particle-Resolved Simulations (PRS), or the Direct Numerical Simulation (DNS) of particle-laden flows, has received constant attention and development, owing to its capability of fully resolving the flow around the particles and not relying on the accuracy of drag models.

Over the past few decades, different PRS methods have been developed. Zhang and Prosperetti~\cite{Zhang2005-tn} developed and Willen and Prosperetti~\cite{Willen2019-rm} employed the PHYSALIS method that assumes flow near spheres to be Stokes flow regardless of the mean flow Reynolds number. As such, PHYSALIS utilizes the analytical solution for Stokes flow around a sphere to simulate particle-laden flows. Glowinski et al.~\cite{Glowinski1999-cx,Glowinski2001-jz} developed a Distributed Lagrange Multiplier (DLM) method that enforces rigid body motion of the fluid in a sphere. Ladd~\cite{Ladd1989-jc} developed the lattice-Boltzmann method (LBM) in which particles are represented by lattice nodes. Another family of methods is the Immersed Boundary Method (IBM) first proposed by Peskin~\cite{Peskin1977-wj,Peskin2002-hm} and then extended to particle-laden flow~\cite{Uhlmann2005-hf,Akiki2016-pq,Biegert2017-ku,Kim2001-ne,Wang2008-fx,Breugem2012-rk, Apte2009-hu, Apte2013-pp, Yu2007-mt}. One challenge associated with moving boundaries is the oscillation of the IBM forces due to (1) the spatial discontinuity in pressure, (2) the temporal discontinuity in the Eulerian velocity at the interface and (3) mass conservation violation~\cite{Huang2019-vz}. Uhlmann~\cite{Uhlmann2005-hf} introduced the concept of direct forcing and reduced the oscillation due to the temporal discontinuity by computing the IBM force at Lagrangian markers and interpolating between an Eulerian and Lagrangian marker with a regularized Dirac delta function. Kempe and Fr\"{o}hlich~\cite{Kempe2012-lp} reduced the minimum stable particle-fluid density ratio from 1.2 to 0.2. Kempe and Fr\"{o}hlich~\cite{Kempe2012-lp} and Wang et al.~\cite{Wang2008-fx} further improved the accuracy of the direct forcing IBM by implementing iterative outer forcing loops to enforce a more accurate no-slip boundary conditions on particle surfaces. Akiki and Balachandar~\cite{Akiki2016-pq} extended IBM to non-uniform grids. Recently, Zhou and Balachandar~\cite{Zhou2021-dg} conducte a theoretical analysis on the optimum number of Lagrangian markers need to achieve most accurate simulations. Yang and Balachandar~\cite{Yang2021-wg} developed a scalable parallel algorithm for IBM with the concept of double binned ghost particle (DBGP). 

For systems with concentrated suspensions, particle-particle interactions are inevitable and collision models are necessary~\cite{Uhlmann2005-hf, Biegert2017-ku, Kempe2012-pl, Yu2016-tx, Xia2020-uw}. In the original direct forcing IBM, Uhlmann~\cite{Uhlmann2005-hf} employed the repulsion potential model proposed by Glowinski et al.~\cite{Glowinski1999-cx}. In this approach, an arbitrary force is added when particles are less than two grid cells apart, a situation not resolvable by IBM. The main purpose of this approach is to prevent particles from contacting one another. Studies that employ this model usually have small volume fractions where particle-particle interactions are not significant~\cite{Kidanemariam2014-qm}. To obtain more accurate particle-particle interactions, collision models have been implemented in the form of lubrication and contact models. Lubrication models account for the forces exerted on particles when they are moving towards and away each other before and after coming into contact, while the contact model is used when the surfaces of the particles touch. For normal contact forces, two popular approaches are the hard- and soft-sphere based models. As discussed by Kempe and Fr\"{o}hlich~~\cite{Kempe2012-lp}, the former cannot model simultaneous collisions which can be significant in concentrated suspensions. In addition, both models usually require the time step size $\Delta{t}$ to be small due to a high material stiffness coefficient $k_n$. To reduce the stiffness, Zaidi et al.~\cite{Zaidi2015-go} chooses a value of $k_n$ that achieves a balance between accuracy and computational cost. Kempe and Fr\"{o}hlich~\cite{Kempe2012-lp}, on the other hand, proposed an algorithm to dynamically optimize the stiffness coefficient $k_n$ and damping coefficient $d_n$ based on the dry restitution coefficient $e_{dry}$ of the material and a calibrated collision time step $T_c$. For tangential contact models, Kempe and Fr\"{o}hlich~~\cite{Kempe2012-lp} designed a model to exactly enforce the no-slip condition between particles. As pointed out by Biegert et al.~\cite{Biegert2017-ku}, this model does not converge to a steady solution for enduring contact. Instead, Luding~\cite{Luding2008-lr} and Thornton et al.~\cite{Thornton2013-uv} proposed and Biegert et al.~\cite{Biegert2017-ku} adopted a spring dashpot model that converged to a steady solution. 

Although there have been significant efforts on developing and improving IBM and collision models, the aforementioned developments are based on the staggered grid formulation. Staggered grids have the advantage of pressure-velocity coupling and satisfying continuity with machine precision. The collocated grid, on the other hand, has the advantage that all variables are located at the same location and the finite-volume approach is straightforward to implement, thus simplifying its use in complex geometries~\cite{Zang1994-ck}. Another significant advantage of the IBM method is its flexibility of implementation in any established Navier-Stokes solver. Lee and Balachandar~\cite{Lee2010-mr,Lee2011-bj} adapted the staggered formulation by Uhlmann~\cite{Uhlmann2005-hf} to a collocated formulation to simulate a single sphere in a shear flow. However, Uhlmann's method does not account for particle collisions and errors due to the explicit formulation of immersed boundary forces~\cite{Kempe2012-lp}. Various authors have developed collocated immersed boundary methods for complex geometries which do not include the effects of interactions between different geometries~\cite{Zhang2021-ai, Mittal2008-ym, Tseng2003-vw,Gilmanov2005-js}. To date, no comprehensive collocated direct forcing IBM with collision models for polydispersed particles has been proposed. In the present work, we build on work by multiple researchers~\cite{Uhlmann2005-hf, Kempe2012-pl, Biegert2017-ku} and propose a simulation framework that combines the advantages of the collocated direct forcing IBM and the collision models for polydisperse particles, resulting in a comparable accuracy with the staggered-grid approach. This work will serve as a building block to couple IBM with existing collocated-grid Navier-Stokes solvers.

The paper is organized as follows. In Section~\ref{sec:original_ibm}, we briefly discuss the original direct-forcing IBM on a staggered grid by Uhlmann~\cite{Uhlmann2005-hf}. In Section~\ref{sec:modified-ibm}, we present the modifications to the original IBM for a collocated grid. Section~\ref{sec:pv-collocate-coupling}-~\ref{sec:triply-bc} focus on the Navier-Stokes solver, Section~\ref{sec:disable_lag}-~\ref{sec:tang_forces} adapt the collision models to a collocated grid and Section~\ref{sec:direct} and~\ref{sec:higher_order} modify the particle solver for coupling of IBM with the collocated-grid Navier-Stokes solver. In Section~\ref{sec:vals}-~\ref{sec:par-par-interaction-val}, simulations are presented to 1) validate the accuracy and computational efficiency of the proposed collocated-grid IBM and 2) determine the tuning parameters that are needed to achieve accurate simulations. In Section~\ref{sec:mono-sim}, monodispersed fluidization simulations are conducted and compared to predictions by existing models.
\section{Methodology}
\subsection{Original direct-forcing Immersed Boundary Method (IBM) on a staggered grid}
\label{sec:original_ibm}
The governing equations are formulated as the modified unsteady Navier-Stokes equation for an incompressible fluid. The IBM force is accounted for with a source term, $\boldsymbol{f}_{\text{IBM}}$, which is added to the Navier-Stokes equation to enforce no-slip boundary conditions on the particle surfaces. With this forcing, the modified unsteady, Navier-Stokes equation for an incompressible fluid is given by
\begin{eqnarray}
&&\pdv{\boldsymbol{u}}{t} + \boldsymbol{u} \vdot \grad{\boldsymbol{u}} = -\grad{P} + \nu_f \laplacian{\boldsymbol{u}} + \boldsymbol{f}_{\text{IBM}}, \label{eq:NS-eqs} 
\end{eqnarray}
subject to continuity,  
\begin{eqnarray}
&&\div{\boldsymbol{u}} = 0,
\label{eq:continuity}
\end{eqnarray}
where $\boldsymbol{u} = \begin{bmatrix}u & v & w \end{bmatrix}^T$ is the fluid velocity vector in Cartesian coordinates, $P = (P_{tot} - \rho_f gz)/\rho_f$ is the perturbation pressure (relative to the hydrostatic pressure $\rho_fgz$, where $g$ is the gravitational acceleration in the z direction) normalized by the fluid density $\rho_f$, $P_{tot}$ is the total pressure, $\nu_f$ is the kinematic viscosity of the fluid and $\boldsymbol{f}_{\text{IBM}} = \begin{bmatrix}f_x & f_y & f_z \end{bmatrix}^T$ is the IBM force vector. The original direct-forcing IBM method proposed by Uhlmann~\cite{Uhlmann2005-hf} eliminates strong oscillations arising from direct interpolation of $\boldsymbol{f}_{\text{IBM}}$ from the neighboring Eulerian cells~\cite{Kim2001-ne,Fadlun2000-jc}. Uhlmann~\cite{Uhlmann2005-hf} proposed to compute the IBM force at the center of Lagrangian markers located at $\boldsymbol{X}_l$ which represent a thin shell of thickness $h$ on the particle surface with volume
\begin{eqnarray}
&&N_l \Delta{V}_l = \frac{4}{3}\pi \left[ \left(\frac{d_p}{2} + \frac{h}{2} \right)^3 - \left(\frac{d_p}{2} - \frac{h}{2} \right)^3 \right],
\end{eqnarray}
where $N_l$ is the total number of Lagrangian markers, $\Delta{V}_l \approx h^3$ is the approximate volume of each Lagrangian marker, $d_p$ is the particle diameter and $h$ is the Eulerian grid spacing, which is isotropic in the three Cartesian coordinate directions (x, y, z) such that $\Delta{x} = \Delta{y} = \Delta{z} = h$. The desired motion of particles at particle surface locations $\boldsymbol{X}_l$ is defined as
\begin{eqnarray}
 && \boldsymbol{u}^d_p \left(\boldsymbol{X}_l \right) = \boldsymbol{u}_p + \boldsymbol{\omega}_p \crossproduct  \left(\boldsymbol{X}_l - \boldsymbol{x}_p \right),
\end{eqnarray}
where $\boldsymbol{u}_p$ and $\boldsymbol{\omega}_p$ are governed by Newton's second law governing linear and angular momentum of a spherical particle
\begin{subequations}
\begin{eqnarray}
&&m_p \dv{\boldsymbol{u}_p}{t} = \oint_{S} \boldsymbol{\tau} \vdot \boldsymbol{n} \ \dd S + V_p(\rho_p-\rho_f)\boldsymbol{g} + \boldsymbol{F}_{c,p}, \label{eq:particle_momemtum} \\
&&I_p \dv{\boldsymbol{\omega}_p}{t} = \oint_{S} \boldsymbol{r}  \crossproduct  \left(\boldsymbol{\tau} \vdot \boldsymbol{n} \right) \ \dd S + \boldsymbol{T}_{c,p},
\end{eqnarray}
\end{subequations}
where $m_p=\rho_p V_p$ is the mass of the particle, $\rho_p$ and $V_p$ are the particle density and volume, $\boldsymbol{\tau}$ is the hydrodynamic stress tensor, $\boldsymbol{n}$ is the outward-pointing normal vector on the particle surface $S$, $\boldsymbol{g}= -g \boldsymbol{e}_z$ is the gravitational acceleration vector in the $\boldsymbol{e}_z$ direction, $I_p$ is the particle moment of inertia, $\boldsymbol{r}$ is the position vector between the particle center $\boldsymbol{x}_p$ and particle surface and $\boldsymbol{F}_{c,p}$ and $\boldsymbol{T}_{c,p}$ are the forces and torques exerted on the particle due to collisions including lubrication and contact forces. Substitution of the Navier-Stokes equation~\ref{eq:NS-eqs} (with the full pressure and gravity terms included) into equation~\ref{eq:particle_momemtum} and employing Gauss's theorem, the particle motion equations are given by
\begin{subequations}
\begin{eqnarray}
\label{eq:par_motion}
&&m_p \dv{\boldsymbol{u}_p}{t} = - \rho_f\int_{\Omega_p} \boldsymbol{f}_{\text{IBM}} \ \dd V + \rho_f\dv{t} \int_{\Omega_p} \boldsymbol{u} \ \dd V + V_p \left(\rho_p - \rho_f \right)\boldsymbol{g} + \boldsymbol{F}_{c,p},  \\
&&I_p \dv{\boldsymbol{\omega}_p}{t} = \rho_f\int_{\Omega_p} \boldsymbol{r}  \crossproduct \boldsymbol{f}_{\text{IBM}} \ \dd V + \rho_f \dv{t} \int_{\Gamma_p} \boldsymbol{r}  \crossproduct \boldsymbol{u} \ \dd V + \boldsymbol{T}_{c,p},
\end{eqnarray}
\end{subequations}
where $\Omega_p$ is the particle volume. Following Uhlmann~\cite{Uhlmann2005-hf}, the rate-of-change terms can be simplified by assuming rigid-body motion such that 
\begin{subequations}
\begin{eqnarray}
&&\dv{t} \int_{\Omega_p} \boldsymbol{u} \ \dd V = V_p \dv{\boldsymbol{u}_p}{t}, \\
&&\dv{t} \int_{\Omega_p} \boldsymbol{r}  \crossproduct \boldsymbol{u} \ \dd V = \frac{I_p}{\rho_p} \dv{\boldsymbol{\omega_p}}{t}.
\end{eqnarray}
\end{subequations}
which gives the governing equations
\begin{subequations}
\begin{eqnarray}
&&m_p \left(1-\frac{\rho_f}{\rho_p} \right) \dv{\boldsymbol{u}_p}{t} = - \rho_f\int_{\Omega_p} \boldsymbol{f}_{\text{IBM}} \ \dd V + V_p \left(\rho_p - \rho_f \right)\boldsymbol{g} + \boldsymbol{F}_{c,p},  \\
&&I_p \left(1-\frac{\rho_f}{\rho_p} \right) \dv{\boldsymbol{\omega}_p}{t} = \rho_f\int_{\Omega_p} \boldsymbol{r}  \crossproduct \boldsymbol{f}_{\text{IBM}} \ \dd V + \boldsymbol{T}_{c,p}.
\end{eqnarray}
\end{subequations}
However, this can lead to a singularity when $\rho_p=\rho_f$, a problem that is addressed in Section~\ref{sec:direct}.

\begin{figure}[ht]
\centerline{
 {\includegraphics[width=\textwidth]{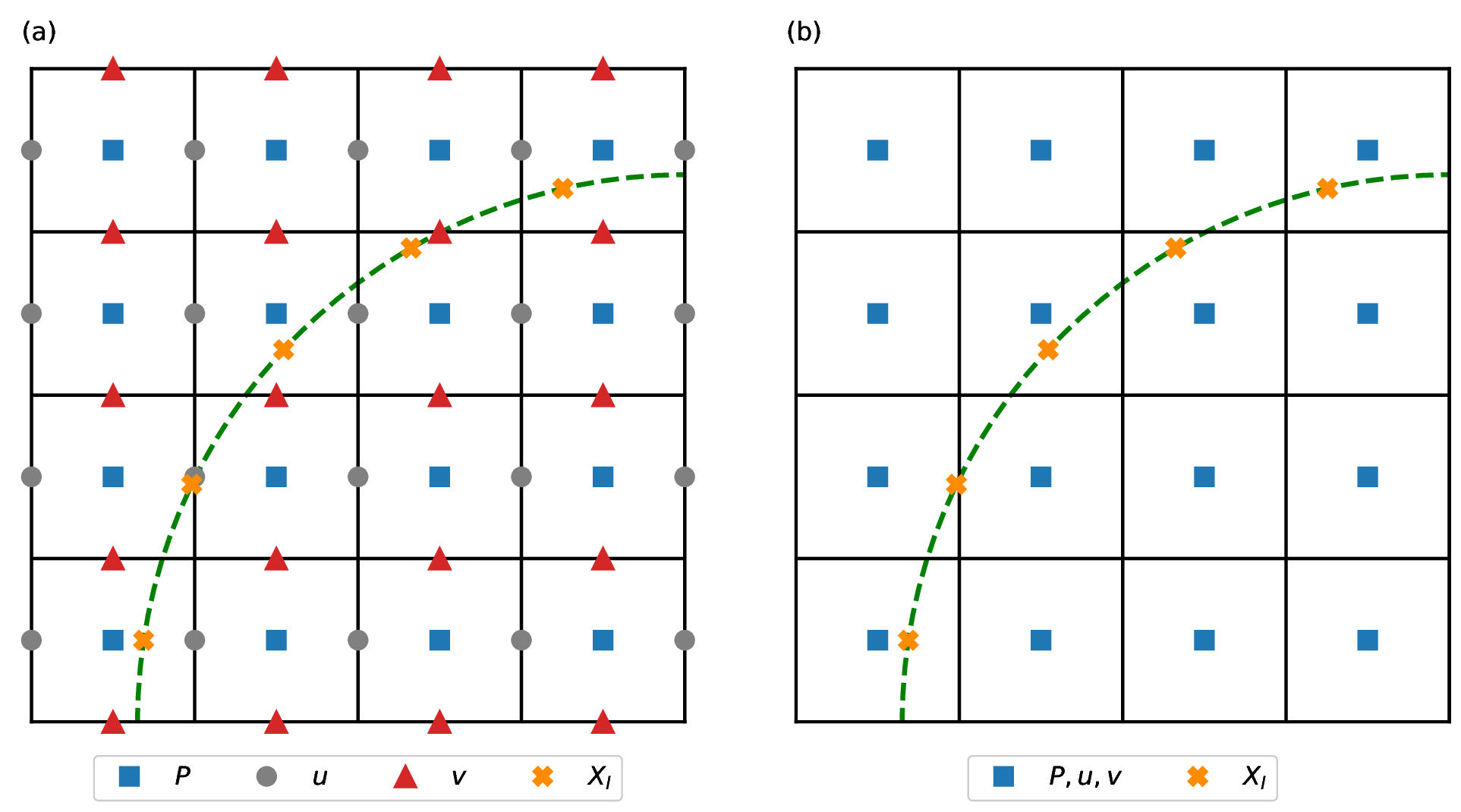}}}
\caption{Variable locations on (a) staggered and (b) collocated grids. Green dashed line represents the particle surface.}
\label{fig:collocated-staggered}
\end{figure}

Figure~\ref{fig:collocated-staggered} shows an example of the Eulerian grid and the Lagrangian markers representing the particle on both staggered and collocated grids. To interpolate between Eulerian and Lagrangian quantities, the one-dimensional kernel in the \emph{x}-direction based on the three-point regularized Dirac delta function $\delta^{3p}_h$ is utilized and defined as
\begin{eqnarray}
\label{eq:3pt-delta}
&&\delta^{3p}_{h,x}  \left(x-X_l \right) = \frac{1}{h} \phi_3 \left(r \right),
\end{eqnarray} 
where $\phi_3 \left(r \right)$ is the one-dimensional, three-point function
\begin{eqnarray}
&&\phi_3 \left(r \right) = \begin{cases}
\frac{1}{6} \left(5 - 3\abs{r} - \sqrt{-3 \left(1-\abs{r} \right)^2+1} \right), & 0.5 \le \abs{r} \le 1.5, \\[10pt]
\frac{1}{3} \left(1 + \sqrt{-3\abs{r}^2 + 1} \right), & \abs{r} \le 0.5, \\[10pt]
0, & \text{otherwise,}
\end{cases}
\label{3pt-stencil}
\end{eqnarray}
and $r =  \left(x-X_l \right) / h$ is the normalized distance from the Lagrangian marker~\cite{Roma1999-cp}. In three dimensions, the three-point regularized Dirac delta function $\delta^{3p}_{h,3D}$ is then given by 
\begin{eqnarray}
&&\delta^{3p}_{h,3D} \left(\boldsymbol{x} - \boldsymbol{X}_l \right) = \delta^{3p}_{h,x}  \left(x-X_l \right) \delta^{3p}_{h,y}  \left(y-Y_l \right) \delta^{3p}_{h,z}  \left(z-Z_l \right).
\end{eqnarray}
This is used to interpolate quantities from the Eulerian grid onto the Lagrangian marker and vice-versa. In what follows, $\delta_{h,3D}$ (without the superscript) implies either the three- or four-point functions which are defined in equation~\ref{3pt-stencil} and equation~\ref{4pt-stencil} respectively. The choice is clarified in the test case.

In Uhlmann~\cite{Uhlmann2005-hf}, the advection term is integrated in time with the
explicit, three-step Runge-Kutta scheme described in Rai and Moin~\cite{Rai1991-dp}. The viscous term is time integrated with the second-order 
implicit Crank-Nicolson scheme to eliminate the associated stability
constraints. The fractional step method by Rai and Moin~\cite{Rai1991-dp} is used to couple the velocity and the pressure terms and enforces continuity. Overall, the fluid-solver with direct-forcing IBM on a uniform staggered grid by Uhlmann~\cite{Uhlmann2005-hf} is given by the following steps: 
\begin{subequations}
\label{eq:basic-fluid-solver}
\begin{enumerate}
    \item Predictor step without direct forcing 
    
    \begin{align}
        \ \ \ \ \ \   \frac{\widetilde{\boldsymbol{u}} - \boldsymbol{u}^{k-1}}{\Delta{t}} &=  2\alpha_k \nu_f \laplacian{\boldsymbol{u}}^{k-1} - 2\alpha_k \grad{P}^{k-1} \nonumber \\ 
        & - \gamma_k  \left(\boldsymbol{u} \dotproduct \grad{\boldsymbol{u}} \right)^{k-1} - \zeta_k  \left(\boldsymbol{u} \dotproduct \grad{\boldsymbol{u}} \right)^{k-2} ,
    \end{align}
    
    \item Project predicted Eulerian grid velocities onto the Lagrangian marker
    \begin{eqnarray}
        && \widetilde{\boldsymbol{U}} \left(\boldsymbol{X}_{l,n} \right) =  \sum_{i,j,k}^{N_i,N_j,N_k} \widetilde{\boldsymbol{u}} \left(\boldsymbol{x}_{i,j,k} \right) \delta^{3p}_{h,3D} \left(\boldsymbol{x_{i,j,k}} - \boldsymbol{X}_{l,n} \right) \Delta{x} \Delta{y} \Delta{z},
    \end{eqnarray}  
    where $\delta^{3p}_{h,3D}$ is the three-point regularized Dirac-delta function (Equation~\ref{eq:3pt-delta}).
    
    \item Determine the Lagrangian marker force to enforce the no-slip condition
    \begin{eqnarray}
        && \boldsymbol{F} \left(\boldsymbol{X}_{l,n} \right) =  \frac{\boldsymbol{u}_p^d \left(\boldsymbol{X}_{l,n} \right) - \widetilde{\boldsymbol{U}} \left(\boldsymbol{X}_{l,n} \right)}{2\alpha_k\Delta{t}},
    \end{eqnarray}  
    
    \item Interpolate the Lagrangian marker force back onto the Eulerian grid
    \begin{eqnarray}
       &&  \boldsymbol{f}_{\text{IBM}} (\boldsymbol{x}_{i,j,k}) =  \sum_{n,l}^{N_p,N_l} \boldsymbol{F} \left(\boldsymbol{X}_{l,n} \right) \delta^{3p}_{h,3D} \left(\boldsymbol{x}_{i,j,k} - \boldsymbol{X}_{l,n} \right) \Delta{V}_{l,n},
    \end{eqnarray}  
    
    \item Predictor step with direct forcing
    \begin{eqnarray}
        && \boldsymbol{u}^* = \widetilde{\boldsymbol{u}} + \Delta{t}  \left(2\alpha_k\boldsymbol{f}_{\text{IBM}} -\alpha_k \nu_f \laplacian{\boldsymbol{u}}^{k-1} + \alpha_k \nu_f \laplacian{\boldsymbol{u}}^* \right),
    \end{eqnarray}  
    
    \item  Pressure Poisson equation to obtain the pseudopressure $\phi$  
    \begin{eqnarray}
        && \laplacian{\phi} = \frac{1}{2\alpha_k\Delta{t}}\div{\boldsymbol{u}^*}, \label{eq:p-poisson}
    \end{eqnarray}  

    \item Corrector step
    \begin{eqnarray}
        && \boldsymbol{u}^k = \boldsymbol{u}^* - 2\alpha_k \Delta{t} \grad{\phi}, \label{eq:p-cor}
    \end{eqnarray}  
    
    \item Compute the real pressure $P$ from the pseudopressure $\phi$
    \begin{eqnarray}
      &&  P^k = P^{k-1} + \phi - \alpha_k\Delta{t}\nu_f \laplacian{\phi}. 
    \end{eqnarray}  
\end{enumerate}
\end{subequations}
In this method, $\alpha_k$, $\gamma_k$ and $\zeta_k$ for $k=1,2,3$ are Runge Kutta coefficients in Rai and Moin~\cite{Rai1991-dp}. To couple the interactions between fluid and markers, $\boldsymbol{u}_p^d \left(\boldsymbol{X}_l \right)$ is computed from the discrete equations that govern the linear and angular particle momentum 
\begin{subequations}
\label{eq:basic-particle}
\begin{eqnarray}
&&\boldsymbol{u}_p^k = \boldsymbol{u}_p^{k-1} + 2\alpha_k \Delta{t} \frac{\rho_p \rho_f}{m_p  \left(\rho_p - \rho_f \right)} \left[-\sum_l^{N_l} \boldsymbol{F} \left(\boldsymbol{X}_l \right) \Delta{V}_l + \frac{\boldsymbol{F}_{c,p}}{\rho_f} \right] + 2\alpha_k \Delta{t} \boldsymbol{g}, \\[7pt] \label{eq:dis-up-ori}
&&\boldsymbol{\omega}_p^k = \boldsymbol{\omega}_p^{k-1} + 2\alpha_k \Delta{t} \frac{\rho_p \rho_f}{I_p  \left(\rho_p - \rho_f \right)} \left[ -\sum_l^{N_l}  \left(\boldsymbol{X}_l - \boldsymbol{x}_p^{k-1} \right) \crossproduct \boldsymbol{F} \left(\boldsymbol{X}_l \right) \Delta{V}_l  + \boldsymbol{T}_{c,p} \right], \\[10pt]
&&\boldsymbol{x}_p^k = \boldsymbol{x}_p^{k-1} + \alpha_k \Delta{t}  \left(\boldsymbol{u}_p^k + \boldsymbol{u}_p^{k-1} \right), \\[10pt]
&&\boldsymbol{u}_p^d \left(\boldsymbol{X}_l \right) = \boldsymbol{u}_p^k + \boldsymbol{\omega}_p^k \crossproduct  \left(\boldsymbol{X}_l - \boldsymbol{x}_p^k \right).
\end{eqnarray}
\end{subequations}
In what follows, Equations~\ref{eq:basic-fluid-solver}(a)-(h) are referred to as the original fluid solver and Equations~\ref{eq:basic-particle}(a)-(d) are referred to as the original particle solver. Thorough validations with both two- and three-dimensional~\cite{Uhlmann2003-ke,Uhlmann2004-fu,Uhlmann2005-hf} cases have been conducted to demonstrate that the original direct-forcing IBM is second-order accurate in both time and space to simulate particle-flow interactions.

\subsection{Finite-volume, Immersed Boundary Method on a collocated grid}
\label{sec:modified-ibm}
\subsubsection{Collocated fluid solver: Pressure-momentum coupling on a collocated grid}
\label{sec:pv-collocate-coupling}
Instead of a finite-difference Navier-Stokes solver on a staggered grid, equation~\ref{eq:NS-eqs} is discretized on a collocated grid using the finite-volume approach. The main disadvantage of collocated grids is the lack of coupling between momentum and pressure when solving the pressure Poisson equation as defined in Equation~\ref{eq:p-poisson}. A collocated grid results in wider stencils to compute the Laplacian term $\laplacian{\phi}$ leading to decoupling between velocity and pressure and only an approximately divergence-free flow~\cite{Zang1994-ck,Rhie1983-ww}. This decoupling results in a ``checkerboard'' pressure field and grid-scale oscillations in the velocity field~\cite{Rhie1983-ww,Ferziger2019-tk}. 

A common method to eliminate the checkerboarding is to use a staggered formulation of the pressure Poisson equation by interpolating quantities from cell centers to faces~\cite{Rhie1983-ww,Zang1994-ck}. Here, we adopt the method of Zang et al.~\cite{Zang1994-ck} in using interpolated face values of the velocity field $\boldsymbol{u}_f$ to solve the pressure Poisson equation and then correcting $\boldsymbol{u}_f$ and $\boldsymbol{u}$ separately with the following steps:
\begin{subequations}

\begin{enumerate}
    \item Interpolate center to face values
    \begin{eqnarray}
        \boldsymbol{u}_f^* = \mathcal{I} \left(\boldsymbol{u}^* \right),
    \end{eqnarray}
    \item  Pressure Poisson equation to obtain the pseudopressure 
    \begin{eqnarray}
        \laplacian{\phi} = \frac{1}{2\alpha_k\Delta{t}}\div{\boldsymbol{u}_f^*},
    \end{eqnarray}  

    \item Corrector step for the cell-centered velocity
    \begin{eqnarray}
        \boldsymbol{u}^k = \boldsymbol{u}^* - 2\alpha_k \Delta{t} \grad{\phi},
    \end{eqnarray}  
    
    \item Corrector step for the face-centered velocity
    \begin{eqnarray}
        \boldsymbol{u}_f^k = \boldsymbol{u}_f^* - 2\alpha_k \Delta{t}  \left(\grad{\phi} \right)_f
    \end{eqnarray}  
\end{enumerate}
\end{subequations}
In step 1, $\mathcal{I} \left(\vdot \right)$ is an interpolation scheme used to obtain $\boldsymbol{u}_f$ from $\boldsymbol{u}$. To ensure overall second-order accuracy, the interpolation scheme used in our modified method is a second-order accurate linear interpolation.
\subsubsection{Collocated fluid solver: Outer forcing}
\label{sec:outer-f}
In the original direct-forcing IBM, no-slip boundary conditions are approximate and incur an error of $\epsilon = \abs{\boldsymbol{u}_p^d \left(\boldsymbol{X}_l \right) - \widetilde{\boldsymbol{U}} \left(\boldsymbol{X}_l \right)}$ due to the explicit formulation of $\boldsymbol{f}_{\text{IBM}}$ in calculating the predictor velocity field $\boldsymbol{u}^*$~\cite{Kempe2012-lp}. Following Kempe and Fr\"{o}hlich~~\cite{Kempe2012-lp}, the difference between the explicit and implicit $\boldsymbol{f}_{\text{IBM}}$ is defined as 
\begin{eqnarray}
\label{eq:f-error}
&& \Delta{\boldsymbol{f}}_{IBM} = -\Delta{t} \left[\alpha_k \nu_f \laplacian \left(\boldsymbol{u}\vdot \grad{\boldsymbol{u}} + \boldsymbol{f}_{\text{IBM}} \right) + 2\alpha_k^2 \nu_f^2 \laplacian \left(\laplacian{\boldsymbol{u}}^k \right) \right].
\end{eqnarray}
Equation~\ref{eq:f-error} implies first-order error with respect to the time-step size $\Delta{t}$. Therefore, sufficiently small $\Delta{t}$ is required to achieve satisfactory results which leads to an increase in computational cost. To remove the limitation of $\Delta{t}$ imposed by this error, outer forcing loops proposed by various researchers~\cite{Kempe2012-lp,Luo2007-lt} are implemented with $n_f$ steps as follows: \\ \\
\begin{subequations}
 \textbf{for $m=1, n_f$}
\begin{enumerate}
    \item Project Eulerian grid velocities onto the Lagrangian marker
    \begin{eqnarray}
       &&\boldsymbol{U}^{m-1} \left(\boldsymbol{X}_{l,n} \right) =  \sum_{i,j,k}^{N_i,N_j,N_k} \boldsymbol{u}^{m-1} \left(\boldsymbol{x}_{i,j,k} \right) \delta_{h,3D} \left(\boldsymbol{x}_{i,j,k} - \boldsymbol{X}_{l,n} \right) h^3
    \end{eqnarray}
    
    \item Determine the Lagrangian marker force to enforce the no-slip condition
    \begin{eqnarray}
        &&\boldsymbol{F}^{m-1} \left(\boldsymbol{X}_{l,n} \right) =  \frac{\boldsymbol{U}^d \left(\boldsymbol{X}_{l,n} \right) - \boldsymbol{U}^{m-1}_{l,n}}{2\alpha_k\Delta{t}}
    \end{eqnarray}
    
    \item Interpolate Lagrangian marker force onto the Eulerian grid
    \begin{eqnarray}
        &&\boldsymbol{f}_{\text{IBM}}^{m-1}(\boldsymbol{x}_{i,j,k}) =  \sum_{n,l}^{N_p,N_l} \boldsymbol{F}_{l,n}^{m-1} \delta_{h,3D} \left(\boldsymbol{x}_{i,j,k} - \boldsymbol{X}_{l,n} \right) \Delta{V}_{l,n}
    \end{eqnarray}
    
    \item Update Eulerian velocity with the computed force
    \begin{eqnarray}
        &&\boldsymbol{u}^{m} = \boldsymbol{u}^{m-1} + 2 \alpha_k \Delta{t} \boldsymbol{f}^{m-1} \label{eq:f-cor-helmotz}\\
        &&\boldsymbol{u}^{m-1} = \boldsymbol{u}^{m} 
    \end{eqnarray}
\end{enumerate}
\textbf{end} 
\end{subequations}
\newpage
In principle, Equation~\ref{eq:f-cor-helmotz} must be solved implicitly with the viscous term which would require a matrix inversion. However, since the Runge-Kutta time step is relatively small, an explicit update is a valid approximation. Although this method alleviates the time-step constraint, the number of outer forcing loops $n_f$ is a tuning parameter. Kempe and Fr\"{o}hlich~\cite{Kempe2012-lp} reported that large $n_f$ will eventually eliminate the error, although $n_f = 3$ represents a good tradeoff between computational cost and accuracy. However, Biegert~\cite{Biegert2018-rh} stated that results can be oscillatory and negatively impact the collision accuracy. Biegert et al.~\cite{Biegert2017-ku} reported $n_f = 1$ is sufficient to obtain accurate results for a single sphere settling in an approximately unbounded (periodic) domain. In Section~\ref{sec:par-par-interaction-val}, we show that $n_f$ has a strong effect on collision model accuracy and hence must be calibrated based on both the fluid-particle interaction and collision models in our collocated method.
\subsubsection{Collocated fluid solver: Three- and four-point Dirac delta function}
\label{sec:3-4-delta}
In the original direct forcing IBM on a staggered grid, the three-point regularized Dirac delta function (equation~\ref{eq:3pt-delta}) is used. However, on a collocated grid, a four-point function that ensures both odd and even grid cells receive the same forcing may be needed to reduce oscillations that occur with the three-point function. To enforce this constraint, the four-point function must satisfy
\begin{eqnarray}
&& \sum_{i \text{ even}} \phi_4 \left(r_i \right) = \sum_{i \text{ odd}} \phi_4 \left(r_i \right) = \frac{1}{2},
\end{eqnarray}
while the three-point function (Equation~\ref{3pt-stencil}) only enforces
\begin{eqnarray}
&& \sum_{i} \phi_3 \left(r_i \right) = 1,  
\end{eqnarray}
where $i$ is an integer indicating one of the cells on the Eulerian grid. Uhlmann~\cite{Uhlmann2004-fu} suggests that a collocated formulation should use the four-point function proposed by Peskin~\cite{Peskin2002-hm} to reduce oscillations that occur with the three-point function. In the four-point function, 
\begin{eqnarray}
\label{eq:4-pt-stencil}
&&\phi_4 \left(r \right) = \begin{cases}
\frac{1}{8} \left(5 + 2r - \sqrt{-7-12r-4r^2} \right), & -2 \le r \le 1 \\[10pt]
\frac{1}{8} \left(3 + 2r + \sqrt{1 - 4r - 4r^2} \right), & -1 \le r \le 0 \\[10pt]
\frac{1}{8} \left(3 - 2r + \sqrt{1 + 4r - 4r^2} \right), & 0 \le r \le 1 \\[10pt]
\frac{1}{8} \left(5 - 2r - \sqrt{-7+12r-4r^2} \right), & 1 \le r \le 2 \\[10pt]
0, & \text{otherwise.}
\end{cases}
\label{4pt-stencil}
\end{eqnarray}
The three-dimensional, four-point regularized Dirac-delta function is then given by 
\begin{eqnarray}
&&\delta^{4p}_{h,3D} \left(\boldsymbol{x} - \boldsymbol{X}_l \right) = \delta^{4p}_{h,x}  \left(x-X_l \right) \delta^{4p}_{h,y}  \left(y-Y_l \right) \delta^{4p}_{h,z}  \left(z-Z_l \right)
\end{eqnarray}
where
\begin{eqnarray}
&&\delta^{4p}_{h,x}  \left(x-X_l \right) = \frac{1}{h} \phi_4 \left(r \right).
\end{eqnarray}
In Section~\ref{sec:fluid-par-validations}, we compare the effects of the three- and four-point regularized Dirac delta functions on the accuracy of direct-forcing IBM in our collocated grid approach.
\subsubsection{Collocated fluid solver: Triply-periodic boundary conditions}
\label{sec:triply-bc}
To simulate particle suspensions in a triply periodic domain, the compatibility condition   ($\int_\Omega \grad{P} \ \dd \Omega = 0$) must be satisfied~\cite{Uhlmann2005-hf}. As demonstrated by H\"{o}fler and Schwarzer~\cite{Hofler2000-wj}, by assuming periodicity and zero net acceleration and decomposing the pressure gradient with 
\begin{eqnarray}
&& \grad{P} = \grad{P}_{aperiodic} + \grad{P}_{periodic},
\end{eqnarray}
where $\grad{P}_{aperiodic}$ and $\grad{P}_{periodic}$ are the aperiodic and periodic components of the pressure gradient, integrating the Navier-Stokes equation~\ref{eq:NS-eqs} over the computational domain volume $\Omega$ gives
\begin{eqnarray}
&& \grad{P}_{aperiodic} = \frac{1}{V_d} \int_\Omega \boldsymbol{f_{\text{IBM}}} \ \dd V   = \overline{\boldsymbol{f}}_{\text{IBM}} ,
\end{eqnarray}
where $\overline{\boldsymbol{f}}_{\text{IBM}}$ is the volume-averaged direct-forcing vector and $\Omega$ is the computational domain with volume $V_d$. To ensure compatibility, $\overline{\boldsymbol{f}}_{\text{IBM}}$ must be subtracted from equation~\ref{eq:NS-eqs} for triply-periodic cases. Since the aperiodic pressure gradient arises from the buoyancy force due to the particles, the force can be computed as the submerged weight of the $N_p$ particles in the system with
\begin{subequations}
\begin{eqnarray}
\label{eq:fe_sw}
 && \overline{\boldsymbol{f}}_{e,sw} = -\sum_{n}^{N_p} \frac{\pi}{6}d_{p,n}^3 \left(\frac{\rho_{p,n} }{\rho_f} - 1 \right) \boldsymbol{g},
\end{eqnarray}
where $d_{p,n}$ and $\rho_{p,n}$ are, respectively, the diameter and density of particle $n$. Alternatively, it can be directly computed discretely as the average direct IBM force over the Eulerain grid cells in the domain with 
\begin{eqnarray}
\label{eq:fe_IBM}
&&\overline{\boldsymbol{f}}_{\text{IBM}} = -\frac{1}{N_iN_jN_k} \sum_{i,j,k}^{N_i,N_j,N_k}  \boldsymbol{f}_{\text{IBM},ijk} ,
\end{eqnarray}
\end{subequations}
which is the average discrete IBM force added to the system due to the presence of particles. Both equations yield the same result but equation~\ref{eq:fe_IBM} incurs slightly more computational cost than equation~\ref{eq:fe_sw} since equation~\ref{eq:fe_sw} is computed only once. Equation~\ref{eq:fe_IBM} is adopted for our proposed method so as to not restrict to spherical particles for possible future extension.

Including the modifications in this section, the modified finite-volume fluid solver for direct-forcing IBM on a collocated grid is given by
\begin{subequations}
\label{eq:fluid-solver}
\begin{enumerate}
    \item Predictor step without direct forcing 
    
    \begin{align}
        \ \ \ \ \ \   \frac{\widetilde{\boldsymbol{u}} - \boldsymbol{u}^{k-1}}{\Delta{t}} &=  2\alpha_k \nu_f \laplacian{\boldsymbol{u}}^{k-1} - 2\alpha_k \grad{P}^{k-1} \nonumber \\ 
        & - \gamma_k  \left(\boldsymbol{u} \dotproduct \grad{\boldsymbol{u}} \right)^{k-1} - \zeta_k  \left(\boldsymbol{u} \dotproduct \grad{\boldsymbol{u}} \right)^{k-2} ,
    \end{align}
    where $\boldsymbol{u} \dotproduct \grad{\boldsymbol{u}}$ and $\laplacian{\boldsymbol{u}}$ are evaluated with second-order accurate finite differences on the collocated grid. \\
    \item  Project the predicted Eulerian grid velocities onto the Lagrangian marker
    \begin{eqnarray}
        &&\widetilde{\boldsymbol{U}} \left(\boldsymbol{X}_{l,n} \right) =  \sum_{i,j,k}^{N_i,N_j,N_k} \widetilde{\boldsymbol{u}} \left(\boldsymbol{x}_{i,j,k} \right) \delta_{h,3D} \left(\boldsymbol{x}_{i,j,k} - \boldsymbol{X}_{l,n} \right) h^3,
    \end{eqnarray}  
    
    \item Determine the Lagrangian marker force to enforce the no-slip condition
    \begin{eqnarray}
        &&\boldsymbol{F} \left(\boldsymbol{X}_{l,n} \right) =  \frac{\boldsymbol{u}_p^d \left(\boldsymbol{X}_{l,n} \right) - \widetilde{\boldsymbol{U}} \left(\boldsymbol{X}_{l,n} \right)}{2\alpha_k\Delta{t}},
    \end{eqnarray}  
    
    \item Interpolate Lagrangian marker force back onto the Eulerian grid
    \begin{eqnarray}
        &&\boldsymbol{f}_{\text{IBM}}\left(\boldsymbol{x}_{i,j,k}\right) =  \sum_{n,l}^{N_p,N_l} \boldsymbol{F} \left(\boldsymbol{X}_{l,n} \right) \delta_{h,3D} \left(\boldsymbol{x}_{i,j,k} - \boldsymbol{X}_{l,n} \right) \Delta{V}_{l,n}.
    \end{eqnarray}  
    The three- and four-point delta functions are compared in Section~\ref{sec:fluid-par-validations}.
    
    \item Predictor step with direct forcing
    \begin{align}
        \label{eq:collo_predictor}
        \ \ \ \ \ \   \frac{\boldsymbol{u}^* - \widetilde{\boldsymbol{u}}}{\Delta{t}} &=  2\alpha_k \left(\boldsymbol{f}_{\text{IBM}}- \overline{\boldsymbol{f}}_{\text{IBM}} \right) -\alpha_k \nu_f \laplacian{\boldsymbol{u}}^{k-1} \nonumber \\
        & + \alpha_k \nu_f \laplacian{\boldsymbol{u}}^{*} +  2\alpha_k \theta \grad{P}^{k-1} , 
    \end{align}  
where $\overline{\boldsymbol{f}}_{\text{IBM}}$ is the submerged weight of the particles from equation~\ref{eq:fe_IBM} and only non-zero for triply-periodic cases.\\
    \item Interpolate the cell-centered velocities onto the faces
    \begin{eqnarray}
        &&\boldsymbol{u}_f^* = \mathcal{I} \left(\boldsymbol{u}^* \right),
    \end{eqnarray}      
    \item Solve the pressure Poisson equation to obtain the pseudopressure $\phi$ 
    \begin{eqnarray}
        &&\laplacian{\phi} = \frac{1}{2\alpha_k\Delta{t}}\div{\boldsymbol{u}_f^*}
    \end{eqnarray}  

    \item Corrector step to obtain the cell- and face-centered quantities
    \begin{eqnarray}
        &&\boldsymbol{u}^k = \boldsymbol{u}^* - 2\alpha_k \Delta{t} \grad{\phi}, \label{eq:p-cor-new}\\
        &&\boldsymbol{u}_f^k = \boldsymbol{u}_f^* - 2\alpha_k \Delta{t} \grad{\phi}_f,
    \end{eqnarray}  
    
    \item  Compute the full pressure P using the pseudopressure $\phi$
    \begin{eqnarray}
        \label{eq:collo_pressure_cor}
        &&P^k =  \left(1-\theta \right) P^{k-1} + \theta \left(\phi - \alpha_k\Delta{t}\nu_f \laplacian{\phi}  \right).
    \end{eqnarray}  
\end{enumerate}
\end{subequations}
Here, $\theta = 0$ for pressure projection and $\theta = 1$ for pressure correction. Compared to the staggered formulation, equation~\ref{eq:fluid-solver}a-\ref{eq:fluid-solver}d are formulated on a collocated grid. Pressure and momentum have been coupled through equation~\ref{eq:collo_predictor}-\ref{eq:collo_pressure_cor}.
The original direct-forcing IBM~\cite{Uhlmann2005-hf} was thoroughly validated with test cases involving fluid-particle interactions. However, in particle suspensions, particle-particle and particle-wall collisions are inevitable. When particles come into contact with one another or a wall, two problems arise. First, the Lagrangian marker cells overlap, rendering the Dirac delta function invalid. Second, the direct-forcing IBM cannot resolve the flow in the small gaps between the particles or particle and wall. To resolve these issues, we adapt the collision models proposed by Biegert et al.~\cite{Biegert2017-ku} for a collocated grid. In this approach, the collision force $\boldsymbol{F}_{c,p}$ and torque $\boldsymbol{T}_{c,p}$ imposed on particle $p$ are given by
\begin{subequations}
\label{f_int}
\begin{eqnarray}
&&\boldsymbol{F}_{c,p} = \sum_{p, q \ne p}^{N_p}  \left(\boldsymbol{F}_{n,pq} + \boldsymbol{F}_{t,pq} \right) + \boldsymbol{F}_{n,pw} + \boldsymbol{F}_{t,pw}, \\
&&\boldsymbol{T}_{c,p} = \sum_{p, q \ne p}^{N_p} R_{pq,cp} \boldsymbol{n}_{pq} \times \boldsymbol{F}_{t,pq} + R_{pw,cp} \boldsymbol{n}_{pw} \times \boldsymbol{F}_{t,pw},
\end{eqnarray}
\end{subequations}
where $\boldsymbol{F}_{n,pq}$ and $\boldsymbol{F}_{t,pq}$ are the normal and tangential
collision forces between particle $p$ and $q$, $\boldsymbol{F}_{n,pw}$ and
$\boldsymbol{F}_{t,pw}$ are the normal and tangential collision forces between
particle $p$ and a wall, $\boldsymbol{n}_{pq}$ is the vector normal to the plane of contact between particles $p$ and $q$, $\boldsymbol{n}_{pw}$ is the vector normal to the wall at the point of contact with particle $q$, $R_{pq,cp} = 0.5\Vert \left( \Vert \boldsymbol{x}_{q} - \boldsymbol{x}_p \Vert + R_p - R_q \right)  \boldsymbol{n}_{pq}\Vert$  is the effective radius between particle $p$
and $q$ and $R_{pw,cp} = \Vert \boldsymbol{x}_w - \boldsymbol{x}_p \Vert$ is the effective radius between particle $p$
and wall $w$. The normal and tangential collisional forces on particle $p$ are defined as 
\begin{subequations}
\begin{align}
\ \ \ \ \ \  \boldsymbol{F}_{n} &= 
\begin{cases}
0, &\zeta_{n} \ge \epsilon_{sep}\\
\boldsymbol{F}_{n,lub},  &0 \le \zeta_{n} \le \epsilon_{sep}, \\
\boldsymbol{F}_{n,con}, &\zeta_{n} \le 0,
\end{cases} \\
\ \ \ \ \ \  \boldsymbol{F}_{t} &=
\begin{cases}
0, &\zeta_{n} \ge \epsilon_{sep}\\
\boldsymbol{F}_{t,con}, &\zeta_{n} \le 0 ,
\end{cases} 
\end{align}
\end{subequations}
where $\boldsymbol{F}_{n,lub}$ is the lubrication force, $\boldsymbol{F}_{n,con}$ is the normal contact force, $\zeta_{n}$ is the separation distance between the surfaces and $\epsilon_{sep} = 1.5h$ for the three-point Dirac-delta function and $\epsilon_{sep} = 2h$ for the four-point Dirac-delta function.

\subsubsection{Collision models: Disabling the Lagrangian markers}
\label{sec:disable_lag}
To resolve issues associated with overlapping Eulerian grid points, we adopt an approach similar to Kempe and Fr\"{o}hlich~\cite{Kempe2012-lp} and Biegert et al.~\cite{Biegert2017-ku} by excluding Lagrangian markers in calculating $\boldsymbol{f}_{\text{IBM}}$ when the distance between a particle and wall satisfies $\zeta_{n,pw} < \epsilon_{sep}$ or the distance between particles $p$ and $q$ satisfies $\zeta_{n,pq} < 2\epsilon_{sep}$. In addition, for particle-particle collisions, instead of excluding Lagrangian markers for both particles as in Kempe and Fr\"{o}hlich~\cite{Kempe2012-lp} and Biegert et al.~\cite{Biegert2017-ku}, we exclude the Lagrangian markers from just one particle, chosen at random. This avoids the scenario where the particle that is surrounded by many neighbors has none of its Lagrangian markers included in the calculation and better enforces no-slip condition on the particle surfaces.

\subsubsection{Collision models: Lubrication model}
\begin{figure}
\centerline{
 \includegraphics[width=\textwidth]{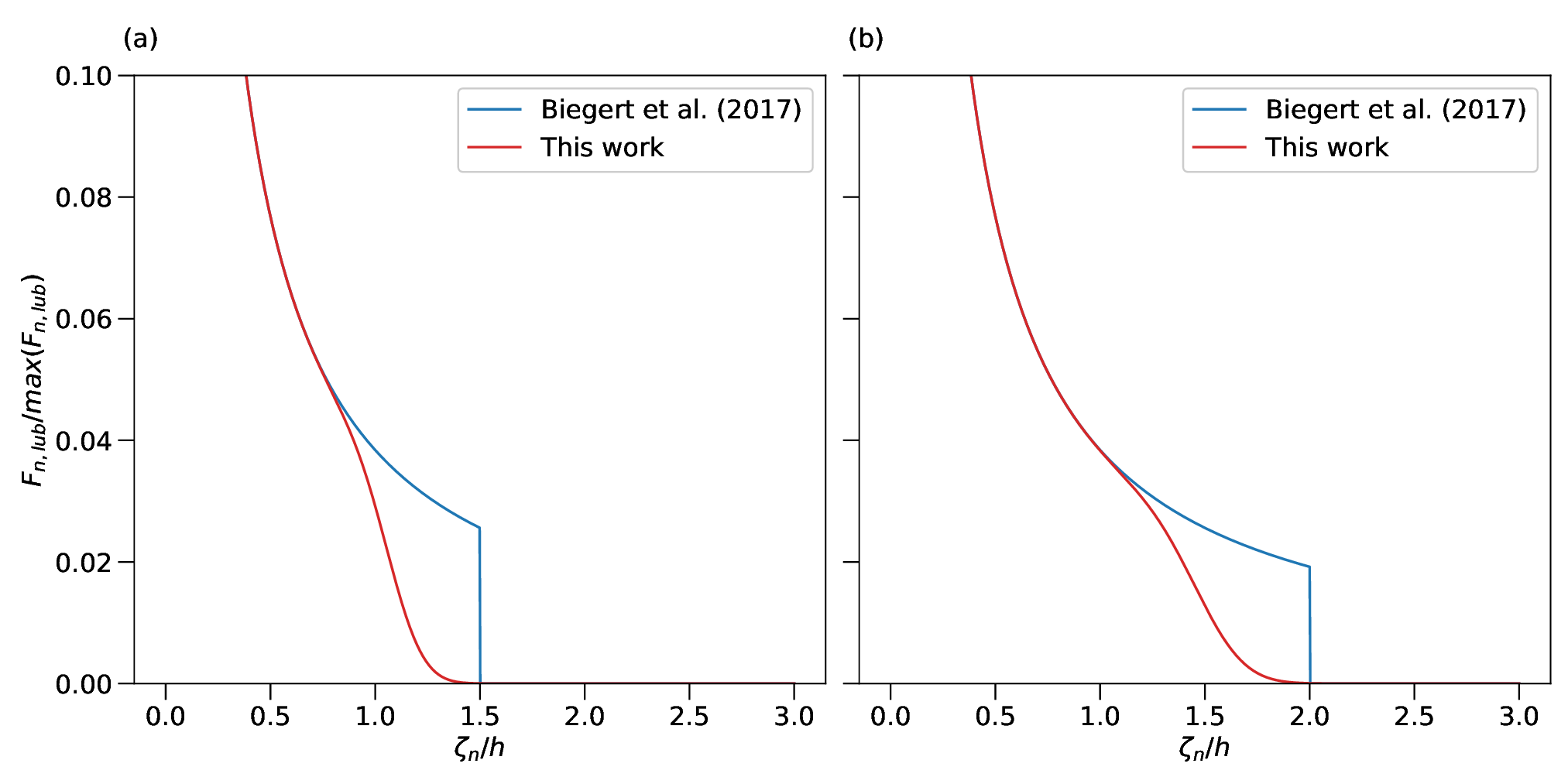}}
\caption{Lubrication force as a function of separation distance between particle $p$ and $q$ for (a) three- and (b) four-point function.}
\label{fig:lub_blend}
\end{figure}

When the separation distance $\zeta_n$ between particles is smaller than the threshold separation distance $ \epsilon_{sep}$, direct-forcing IBM can no longer resolve the flow. Therefore, a lubrication model is used to model the force exerted by the fluid on the particles. We adopt the analytical lubrication model by Cox and Brenner~\cite{Cox1967-wn} and modified by Biegert et al.~\cite{Biegert2017-ku}, in which the lubrication force is given by
\begin{eqnarray}
&&\boldsymbol{F}_{n,lub} = -\frac{6\pi \rho_f \nu_f R_{\text{eff}}^2\psi \left(\zeta_n \right)}{\max \left({\zeta_{\min}, \zeta_n} \right)} \boldsymbol{g}_{n,cp},
\end{eqnarray}
where $R_{\text{eff}} = R_pR_q/(R_p+R_q)$ is the effective radius that is defined based on the particles p and q, $\boldsymbol{g}_{n,cp}$ is the normal component of the relative velocity of the particle surface at the contact point and $\zeta_{\min}$ is the minimum separation distance to prevent a singularity as $\zeta_n \rightarrow 0$~\cite{Biegert2017-ku}. In Biegert et al.~\cite{Biegert2017-ku}, $\psi \left(\zeta_n \right) = \zeta_n$ was used. However, this formulation results in a discontinuity at the interface where $\boldsymbol{F}_{n,lub} \ne 0$ (Figure~\ref{fig:lub_blend}). To avoid the discontinuity, we introduce a function $\psi \left(\zeta_n \right) = 0.5 \erf  \left(6-C\max \left({\zeta_{\min}, \zeta_n} \right)/h \right) + 0.5$ to enforce a continuous force when particles come close to one another where $C = 5.5$ or $4$ for three and four-point function respectively. 

\subsubsection{Collision models: Normal contact model}
To account for the normal contact force during collisions, we implemented the approach by Biegert et al.~\cite{Biegert2017-ku} who employ the adaptive collision time model (ACTM) proposed by Kempe and Fr\"{o}hlich~~\cite{Kempe2012-pl}. The idea behind ACTM is to derive an optimized stiffness coefficient $k_n$ and damping coefficient $d_n$ to achieve the desired dry restitution coefficient $e_{\text{dry}}$ over a collision time $T_c$, as discussed below. The normal contact force $F_{n,con}$ is defined as 
\begin{eqnarray}
\label{eq:hertz-contact}
&&\boldsymbol{F}_{n,con} = -k_n \abs{\zeta_n}^{3/2}\boldsymbol{n} - d_n \boldsymbol{g}_{n,cp},
\end{eqnarray} 
where $k_n$ and $d_n$ are the optimized stiffness and damping coefficients. Equation~\ref{eq:hertz-contact} describes the collision model based on the contact theory of Hertz~\cite{Kempe2012-pl}. In the past, $k_n$ and $d_n$ were typically chosen based on material properties. However, the chosen $k_n$ is usually large, making the equations very stiff. Therefore, the time step size $\Delta{t}$ must be based on the collision model in which the collision time $T_c$ is much smaller than $\Delta{t}$ for the flow calculation. To overcome this problem, instead of fixing $k_n$ and $d_n$, Kempe and Fr\"{o}hlich~~\cite{Kempe2012-pl} proposed to fix $e_{dry}$ and $T_c$ by dynamically optimizing $k_n$ and $d_n$. $e_{dry}$ is a parameter based on the material property of the particle and $T_c$ is a tuning parameter. Large $T_c$ will lead to extensive overlap between colliding particles, making collisions unrealistic, while small $T_c$ will increase the stiffness of the equation. Based on the studies conducted by Kempe and Fr\"{o}hlich~~\cite{Kempe2012-pl}, $T_c = 10\Delta{t}$ represents a good balance between accuracy and computational cost. To obtain $k_n$ and $d_n$, the nonlinear ordinary differential equations representing the interparticle spacing are solved, such that
\begin{subequations}
\label{eq:ode}
\begin{eqnarray}
&&m_{\text{eff}} \dv[2]{\zeta_n}{t} + d_n \dv{\zeta_n}{t} + k_n \zeta_n^{3/2} = 0,\\
&&\dv{\zeta_n}{t} = -\boldsymbol{g}_{n,cp} \vdot \boldsymbol{n},
\end{eqnarray}
\end{subequations}
where $m_{\text{eff}} = m_p m_q / (m_p + m_q)$ is the effective mass accounting for polydisperse particles, and the equations are subject to two conditions at $t=T_c$, $\zeta_n \left(T_c \right) = 0$ and $g_{n,cp} \left(T_c \right) = e_{dry} \boldsymbol{g}_{n,cp} \vdot \boldsymbol{n}$. Kempe and Fr\"{o}hlich~~\cite{Kempe2012-pl} used the Newton-Ralphson method to obtain a solution to equation~\ref{eq:ode}(a) while Ray et al.~\cite{Ray2015-xw} developed an analytical approach. The method of Ray et al.~\cite{Ray2015-xw} incurs less than 1.3\% error in the rebound velocity when $e_{dry} > 0.7$. In our simulations, since $e_{dry} > 0.9$ is typically used, we adopt their approach due to its low computational cost and ease of implementation. 

A potential issue with ACTM as pointed out by Biegert et al.~\cite{Biegert2017-ku} is large $k_n$ for weak collisions. As the impact velocity between collisions $u_{in} \rightarrow 0$, the Stokes number defined as 
\begin{eqnarray}
\label{eq:stokes_num}
&&St = \frac{u_{in}\rho_p d_p}{9 \rho_f \nu_f}
\end{eqnarray}  
also approaches 0, making $k_n \rightarrow \infty$. Both Kempe and Fr\"{o}hlich~\cite{Kempe2012-pl} and Biegert et al.~\cite{Biegert2017-ku} introduce a critical $St_{crit}$ where $u_{in}$ is based on a prescribed $St_{crit}$. In our approach, we set $St_{crit} = 5$ following Biegert et al.~\cite{Biegert2017-ku}. In addition, Biegert et al.~\cite{Biegert2017-ku} also introduces a threshold $k_{n,grav}$ to prevent extensive overlap when $u_{in,crit}$ is large relative to the particle size and relevant time scales for a low Reynolds number flow. Therefore, $k_n$ is defined as 
\begin{eqnarray}
&& k_n = 
\begin{cases}
k_{n,ACTM}, &u_{in} > u_{in,crit} \\[10pt]
\max \left(k_{n,crit}, k_{n,grav} \right), &u_{in} \le u_{in,crit} 
\end{cases}
\end{eqnarray}
where 
\begin{eqnarray}
&&k_{n,crit}= \frac{m_{\text{eff}}}{\sqrt{u_{in,crit}t_*^5}}
\end{eqnarray}
and 
\begin{eqnarray}
&&k_{n,grav} = \max \left(m_p g  \left(\epsilon d_p/2 \right)^{-3/2} , m_q g  \left(\epsilon d_q/2 \right)^{-3/2}  \right),
\end{eqnarray}
where $\epsilon = 10^{-3}$.
\subsubsection{Collision models: Tangential contact model}
\label{sec:tang_forces}
To account for the tangential contact force during collisions, we follow the approach by Biegert et al.~\cite{Biegert2017-ku} who employ the model in a review paper by Thornton et al.~\cite{Thornton2011-dg}. This model uses a spring-dashpot model in which $\boldsymbol{F}_{t,con}$ is defined as
\begin{eqnarray}
&& \boldsymbol{F}_{t,con} = \min \left(\norm{\boldsymbol{F}_{t,dp}}, \norm{\mu_{fri}\boldsymbol{F}_n} \right)\boldsymbol{t},
\end{eqnarray} 
where $\mu_{fri}$ is the coefficient of friction between two surfaces and $\boldsymbol{t} = \boldsymbol{F}_{t,dp} / \norm{\boldsymbol{F}_{t,dp}}$ is the direction vector of the tangential force. $\boldsymbol{F}_{t,dp}$ is defined as 
\begin{eqnarray}
&&\boldsymbol{F}_{t,dp} = -k_t \boldsymbol{\zeta}_t - d_t \boldsymbol{g}_{t,cp},
\end{eqnarray}
where $k_t$ and $d_t$ are the stiffness and damping coefficients, $\boldsymbol{g}_{t,cp}$ is the tangential velocity relative to the surface of contact and $\boldsymbol{\zeta}_t$ is the time-cumulative tangential spring displacement defined as 
\begin{eqnarray}
&& \boldsymbol{\zeta}_t = \int_{t_0}^{t} \boldsymbol{g}_{t,cp} \left(\tau \right) \ \dd \tau,
\end{eqnarray}
where $t_0$ is the impact time. An approach similar to ACTM is adopted for $k_t$ and $d_t$ which are calculated dynamically at each time step. The tangential stiffness coefficient is defined as 
\begin{eqnarray}
&& k_t = \frac{2 \left(1-\nu_{poi} \right)}{2-\nu_{poi}},
\end{eqnarray}
and the tangential damping coefficient is defined as 
\begin{eqnarray}
&& d_t = 2\sqrt{m_{\text{eff}}k_t} \frac{-\ln{e_{dry}}}{\sqrt{\pi^2 +  \left(\ln{e_{dry}}^2 \right)}},
\end{eqnarray}
where $\nu_{poi}$ is Poisson's ratio of the particle material. When compared to the model proposed by Kempe and Fr\"{o}hlich~~\cite{Kempe2012-lp} that enforces slip conditions, this model allows the particles to interact smoothly and stably. To differentiate between rolling/sticking and sliding motions, we adopted the methods by Biegert et al.~\cite{Biegert2017-ku} who employed the formulation of Luding~\cite{Luding2008-lr} in which $\mu_{fri} = \mu_{s}$ when particles are sticking  ($\norm{\boldsymbol{F}_{t,dp}} < \norm{\mu_{fri}\boldsymbol{F}_n}$) and $\mu_{fri} = \mu_{k}$ when slipping occurs ($\norm{\boldsymbol{F}_{t,dp}} > \norm{\mu_{fri}\boldsymbol{F}_n}$).
\subsubsection{Collocated particle solver: Direct computation of fluid inertia within the particle}
\label{sec:direct}
In the original direct-forcing IBM, Uhlmann~\cite{Uhlmann2005-hf} used a rigid body approximation to calculate the rate-of-change term that describes the effect of fluid inertia within the particle. As a result, Equation~\ref{eq:dis-up-ori} has a singularity at $\rho_p/\rho_f = 1$ when $\rho_p - \rho_f = 0$, and the method becomes unstable when $\rho_p/\rho_f < 1.2$. To resolve this issue, Kempe and Fr\"{o}hlich~~\cite{Kempe2012-lp} adopted a level-set approximation to compute the rate-of-change term directly which eliminates both the singularity and improves the stability related to fluid-particle coupling when $\rho_p/\rho_f > 0.2$. Using second-order midpoint quadrature rules, the integrals in the rate-of-change terms are approximated with 
\begin{subequations}
\begin{eqnarray}
&&\int_{\Omega_p} \boldsymbol{u} \ \dd \boldsymbol{V} \approx \sum_{i,j,k}^{N_i,N_j,N_k} \alpha_{i,j,k} \Delta{\Omega}_{i,j,k}  \boldsymbol{u}_{i,j,k}, \\
&&\int_{\Omega_p} \boldsymbol{r} \crossproduct \boldsymbol{u} \ \dd \boldsymbol{V} \approx \sum_{i,j,k}^{N_i,N_j,N_k} \alpha_{i,j,k} \Delta{\Omega}_{i,j,k}  \left( \boldsymbol{r} \crossproduct \boldsymbol{u}  \right),
\end{eqnarray}
\end{subequations}
where $\Delta{\Omega}$ is the volume of a grid cell and $\alpha_{i,j,k}$ is the volume fraction of the cell with indices $i,j,k$ and is defined as 
\begin{eqnarray}
&& \alpha_{i,j,k} = \frac{\Delta{\Omega}_{i,j,k}^{p}}{\Delta{\Omega}_{i,j,k}},
\end{eqnarray}
where  $\Delta{\Omega}_{i,j,k}^{p}$ is the volume of the cell occupied by particle $p$. $\alpha_{i,j,k}$ can be calculated with the level-set approximation
\begin{eqnarray}
&& \alpha_{i,j,k} = \frac{\sum_{m}^8 - \phi_m H \left(-\phi_m \right)}{\sum_{m}^8  \abs{\phi_m}},
\end{eqnarray}
where $H \left(\vdot \right)$ is the Heaviside function
\begin{eqnarray}
&&H \left(-\phi \right) =
\begin{cases}
1, & \phi \le 0, \\
0, & \phi > 0,
\end{cases}
\end{eqnarray}
and $\phi$ is the distance from the corner of each grid cell to the center of the particle and $m= 1,2,\cdots,8$ is the index of the eight corners of a Cartesian grid cell.  

\subsubsection{Collocated particle solver: High-order time integration scheme and sub-stepping}
\label{sec:higher_order}
In the original direct-forcing IBM, the linear and angular momentum equations for the particle motion are integrated in time with the first-order forward Euler scheme or the second-order Crank-Nicolson scheme. Various researchers~\cite{Uhlmann2005-hf,Kempe2012-lp} have shown that these discretizations can produce accurate results of fluid-particle interactions. However, Biegert et al.~\cite{Biegert2017-ku} demonstrated that lower-order schemes do not produce accurate particle rebound velocities. As a result, a collision time of $T_c = 1000\Delta{t}$ is required to reduce the error in the rebound velocity to 0.1\%. Therefore, we followed the approach by Biegert et al.~\cite{Biegert2017-ku} by adopting a higher-order time-stepping scheme with predictor-corrector steps. 

Another issue identified by many researchers~\cite{Biegert2017-ku,Costa2015-ze,Kidanemariam2014-qm} is that a time-step size that accurately resolves fluid-particle interactions may fail to  resolve the lubrication force. To overcome this issue, Costa et al.~\cite{Costa2015-ze} and Biegert et al.~\cite{Biegert2017-ku} proposed sub-iterations for the particle motion solver. Costa et al.~\cite{Costa2015-ze} conducted a total of 50 sub-iteration by using $\Delta{t}_{sub} = \Delta{t} / 50$ while Biegert et al.~\cite{Biegert2017-ku} conducted a total of 15 sub-iterations with $\Delta{t}_{sub} = \Delta{t} / 15$. With the approach by Biegert et al.~\cite{Biegert2017-ku}, each sub-step employs a three-step Runge-Kutta scheme, resulting in a total of 45 iterations. The Runge-Kutta sub-step $k$ of the improved form of the original update given in equations~\ref{eq:basic-particle}(a)-(d) is given by 
\begin{subequations}
\begin{eqnarray}
&&\frac{\widetilde{\boldsymbol{u}}_p - \boldsymbol{u}_p^{k-1}}{\Delta{t}} =   \frac{1}{m_p} \left( 2\alpha_k \boldsymbol{F}_{h,p}^k +  \gamma_k \boldsymbol{F}_{c,p} \left(\boldsymbol{x}_p^{k-1},\boldsymbol{u}_p^{k-1} \right) + \zeta_k \boldsymbol{F}_{c,p} \left(\boldsymbol{x}_p^{k-2},\boldsymbol{u}_p^{k-2} \right) \right) + 2\alpha_k \boldsymbol{g}^\prime, \\
&&\frac{\widetilde{\boldsymbol{\omega}}_p - \boldsymbol{\omega}_p^{k-1}}{\Delta{t}} =   \frac{1}{I_p} \left( 2\alpha_k \boldsymbol{T}_{h,p}^k +  \gamma_k \boldsymbol{T}_{c,p} \left(\boldsymbol{x}_p^{k-1},\boldsymbol{u}_p^{k-1} \right) + \zeta_k \boldsymbol{T}_{c,p} \left(\boldsymbol{x}_p^{k-2},\boldsymbol{u}_p^{k-2} \right) \right), \\
&&\frac{\widetilde{\boldsymbol{x}}_p - \boldsymbol{x}_p^{k-1}}{\Delta{t}} =  \alpha_k  \left(\widetilde{\boldsymbol{u}}_p + \boldsymbol{u}_p^{k-1} \right), \\
&&\frac{\boldsymbol{u}^k_p - \boldsymbol{u}_p^{k-1}}{\Delta{t}} =   \frac{1}{m_p} \left( 2\alpha_k \boldsymbol{F}_{h,p}^k +  \gamma_k \boldsymbol{F}_{c,p} \left(\widetilde{\boldsymbol{x}}_p,\widetilde{\boldsymbol{u}}_p  \right) + \zeta_k \boldsymbol{F}_{c,p} \left(\boldsymbol{x}_p^{k-1},\boldsymbol{u}_p^{k-1} \right) \right) + 2\alpha_k \boldsymbol{g}^\prime, \\
&&\frac{\boldsymbol{\omega}^k_p - \boldsymbol{\omega}_p^{k-1}}{\Delta{t}} =   \frac{1}{I_p} \left( 2\alpha_k \boldsymbol{T}_{h,p}^k +  \gamma_k \boldsymbol{T}_{c,p} \left(\widetilde{\boldsymbol{x}}_p,\widetilde{\boldsymbol{u}}_p \right) + \zeta_k \boldsymbol{T}_{c,p} \left(\boldsymbol{x}_p^{k-1},\boldsymbol{u}_p^{k-1} \right) \right), \\
&&\frac{\boldsymbol{x}^k_p - \boldsymbol{x}_p^{k-1}}{\Delta{t}} =  \alpha_k  \left(\boldsymbol{u}^k_p + \boldsymbol{u}_p^{k-1} \right),
\end{eqnarray}
\end{subequations}
where 
\begin{eqnarray}
&& \boldsymbol{F}_{h,p}^k = 
\begin{cases}
-\rho_f \sum_l^{N_l} \boldsymbol{F}_l \left(\boldsymbol{X}_l \right) \Delta{V}_l + \rho_f \left[\dv{t} \int_{\Gamma_p} \boldsymbol{u} \ \dd \boldsymbol{x} \right]^k, & \max \left(St \right) > St_{crit} \\[10pt]
0, & \max \left(St \right) \le St_{crit}
\end{cases}
\\
&& \boldsymbol{g}^\prime =  \left(1 - \rho_f/\rho_g \right) \boldsymbol{g}, \\[10pt]
&& \boldsymbol{T}^k_{h,p} = 
\begin{cases}
-\rho_f \sum_l^{N_l}  \left(\boldsymbol{X}_l - \boldsymbol{x}_p^{k-1} \right) \crossproduct \boldsymbol{F} \left(\boldsymbol{X}_l \right) \Delta{V}_l + \rho_f \left[\dv{t} \int_{\Gamma_p} \boldsymbol{r}  \crossproduct \boldsymbol{u} \ \dd \boldsymbol{x} \right]^k, & \max \left(St \right) > St_{crit} \\[10pt]
0. & \max \left(St \right) \le St_{crit}
\end{cases}
\end{eqnarray}

\section{Results and discussions}

\subsection{Verification with analytical Taylor Green vortices}
\label{sec:vals}
Following Uhlmann~\cite{Uhlmann2005-hf}, to verify the accuracy of the fluid solver of the finite-volume-based IBM on a collocated grid, we computed the errors associated with computing the flow around a particle located in a flow field given by Taylor Green decaying vortices, for which the analytical solution is given by
\begin{subequations}
\label{eq:taylor}
\begin{eqnarray}
&& u \left(x,y,t \right) = \sin \left(k_xx \right)\cos \left(k_yy \right)\exp \left(- \left(k_x^2+k_y^2 \right)\nu_f t \right),\\
&& v \left(x,y,t \right) = -\frac{k_x}{k_y}\cos \left(k_xx \right)\sin \left(k_yy \right)\exp \left(- \left(k_x^2+k_y^2 \right)\nu_f t\right), \\
&& P \left(x,y,t \right) = \frac{1}{2}\left(-\sin^2 \left(k_xx \right) + \frac{k_x^2}{k_y^2}\cos^2 \left(k_yy \right) \right)\exp \left(-2 \left(k_x^2+k_y^2 \right)\nu_f t \right),
\end{eqnarray}
\end{subequations}
where $k_x = k_y = \pi$~m$^{-1}$ is assumed. The flow is initialized at time $t=0$ with equations~\ref{eq:taylor} and a two-dimensional circular disk with a diameter $d_p = 2$~m and particle-fluid density ratio $s= \rho_p/\rho_f = 1$ is located at the center of the computational domain of size $1.5 d_p \times 1.5 d_p$. The kinematic viscosity $\nu_f = 0.2$~m$^2$~s$^{-1}$ is used. The simulation time is 0.5~s with a time step size $\Delta{t} = 0.001$~s. The desired velocity at the disk surface $\boldsymbol{u}_p^d$ is computed with equations~\ref{eq:taylor}, and hence the desired velocity at the Lagrangian markers is the same as the exact Eulerian velocity at those points. Therefore, the particle should not translate or rotate because there is no viscous stress on the particle surface. An accurate IBM method therefore should give the velocity and pressure fields given by equations~\ref{eq:taylor} since these represent the fluid motion in the absence of a particle. 

\begin{figure}
\centerline{
 \includegraphics[width=\textwidth]{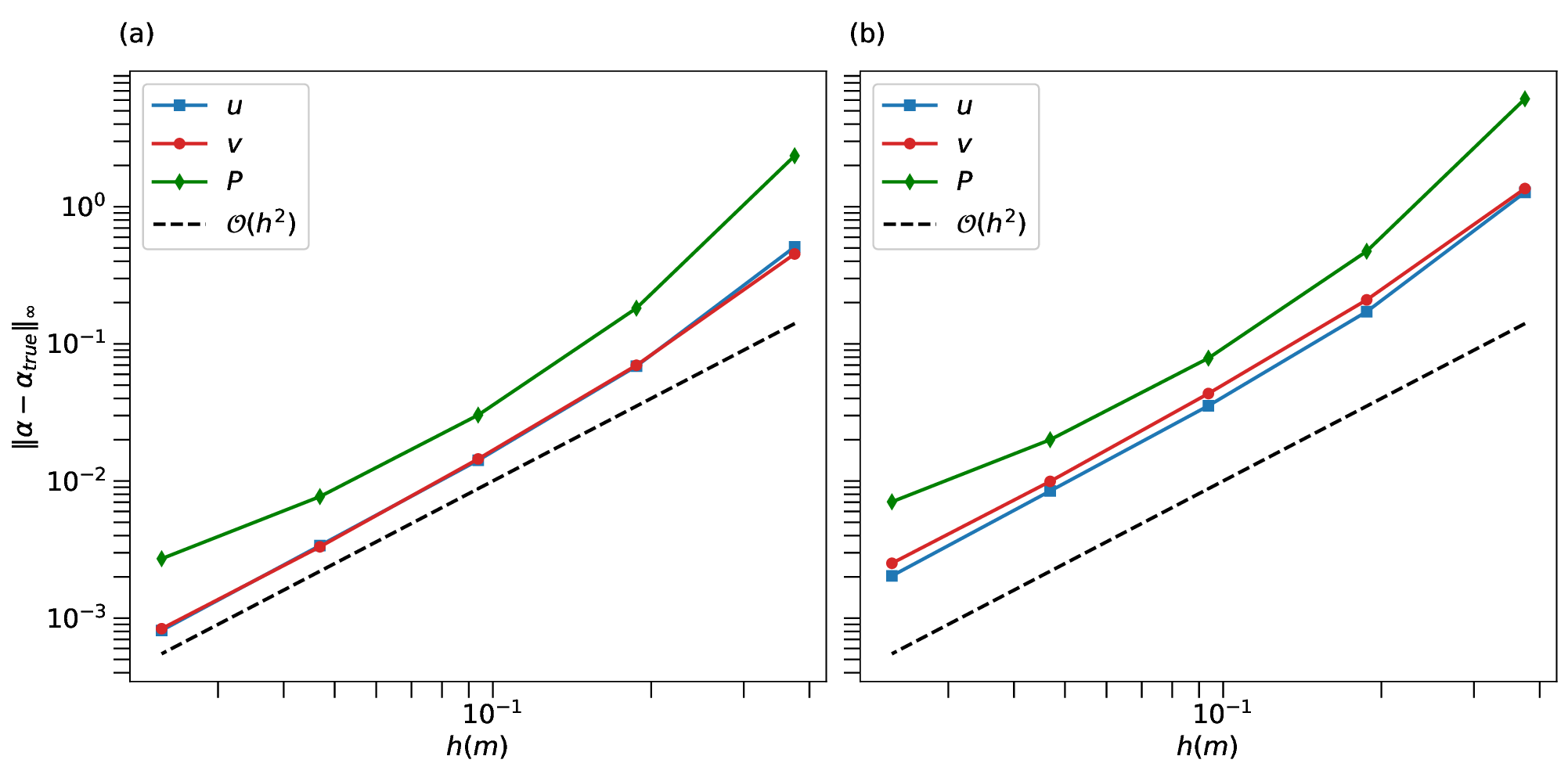}}
\caption{Error of velocity and pressure field for simulating two-dimensional decaying vortices. The error is shown as a function of mesh resolution $h$ (a) without and (b) with a particle.}
\label{fig:verification}
\end{figure}

Figure~\ref{fig:verification} shows the computed errors in the velocity and pressure fields as a function of the grid resolution. The error is given by the $L_\infty$ norm as 
\begin{eqnarray}
\label{eq:linf_err}
&&\text{error}_\infty = \Vert \alpha - \alpha_{true} \Vert_\infty,
\end{eqnarray}
where $\alpha$ and $\alpha_{true}$ are the quantities of interest from simulations and equation~\ref{eq:taylor}, respectively. In this verification analysis, the pressure projection method ($\theta = 0$ in equation~\ref{eq:fluid-solver}) and three-point Dirac delta function are used. Similar trends were observed with other combinations (i.e., pressure projection with four-point Dirac delta function). By comparing to the reference line (dashed-line), second-order convergence in both velocity and pressure fields was observed. For cases with and without the particle, second-order convergence was also observed (figure~\ref{fig:verification}(b) vs. figure~\ref{fig:verification}(a)), showing that the inclusion of interpolation using the discrete delta function does not have an impact on the overall accuracy of the fluid solver. 

To further verify the accuracy of the discrete delta function, we used the same configuration, except that we enforced a no-slip boundary condition on the particle surface such that $\boldsymbol{u}^d_p=0$. Since exact solutions are not available due to the presence of the particle, we use the simulation with the highest resolution ($d_p/h=683)$ as a reference solution, where the highest resolution used is much finer than the typical resolution used for IBM methods($\sim d_p/h=20$). Both $L_2$ and $L_\infty$ norms were computed with 
\begin{subequations}
\begin{eqnarray}
\label{eq:l2_linfty_err}
&&\text{error}_2 = \Vert \alpha - \alpha_{d_p/h=683} \Vert_2,  \\
&&\text{error}_\infty = \Vert \alpha - \alpha_{d_p/h=683} \Vert_\infty.
\end{eqnarray}
\end{subequations}
By comparing to the reference lines (dashed-line for $\mathcal{O}(h^2)$ and dashed-dotted-line for $\mathcal{O}(h)$), approximately first-order convergence in both velocity and pressure fields was observed. Overall, the finite-volume-based Immersed Boundary Method on a collocated grid is approximately first-order accurate. This is consistent with other similar approaches~\cite{Breugem2012-rk}. 

\begin{figure}
\centerline{
 \includegraphics[width=\textwidth]{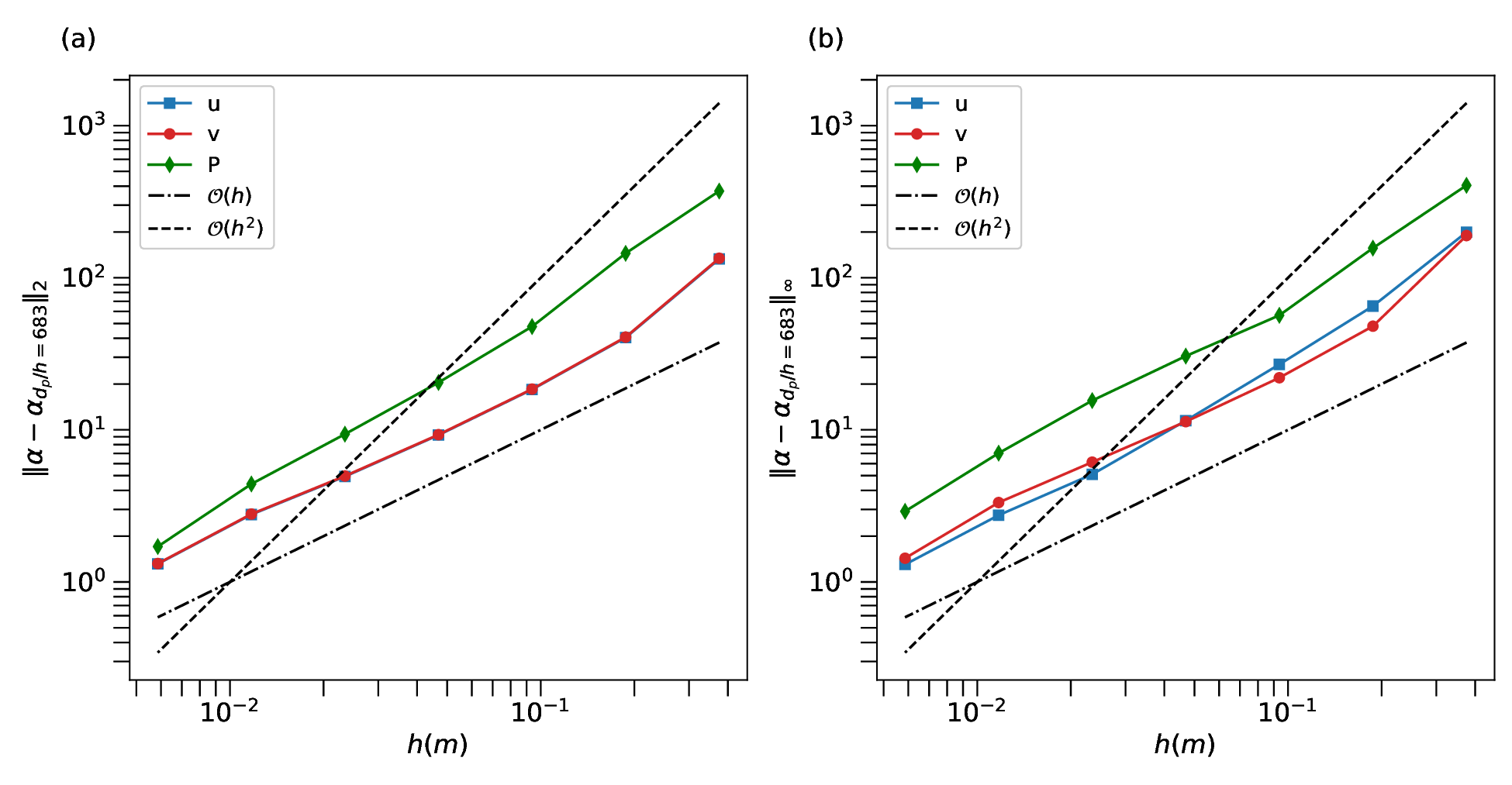}}
\caption{Error of velocity and pressure when simulating two-dimensional decaying vortices with a no-slip condition on a particle. Errors are shown as functions of mesh resolution $h$ for (a) $L_2$  and (b) $L_\infty$ norms.}
\label{fig:verification_delta}
\end{figure}

\subsection{Fluid-particle interactions}
\label{sec:fluid-par-validations}
We simulated the settling of a single particle to validate the accuracy of the fluid-particle interactions using our collocated direct-forcing IBM approach. The primary parameter of interest is the terminal Reynolds number
\begin{eqnarray}
\label{eq:Retinf}
&& Re_{t,\infty} = \frac{w_{t,\infty} d_p}{\nu_f},
\end{eqnarray} 
where $w_{t,\infty}$ is the terminal velocity of a single particle in an approximately unbounded (periodic) domain. In the results, the settling velocity and time are normalized by $w_{ref} = \sqrt{g d_p}$ and $t_{ref} = \sqrt{d_p/g}$, respectively, where $g=9.81$~m~s$^{-2}$. For all simulations in this section, the time step is determined based on a maximum Courant number $C_{\max} = u_{\max}\Delta{t}/h = 0.4$, where $u_{\max}$ is the maximum magnitude of the fluid velocity vector over the course of the simulation. We also assume $n_f = 2$ outer forcing loops (Section~\ref{sec:outer-f}), the three-point Dirac delta function (equation~\ref{3pt-stencil}), and the pressure projection scheme ($\theta = 0$ in equation~\ref{eq:fluid-solver}). Results of a particle settling onto a bottom wall are compared to the experiments of Ten Cate et al.~\cite{Ten_Cate2002-dg}, while results of a particle settling in an approximately unbounded (periodic) domain are compared to experiments of Mordant and Pinton~\cite{Mordant2000-lz}. Simulation parameters and setup are summarized in Table~\ref{tab:fluid-particle-val}. 

\begin{landscape}
\begin{table}[ht!]
\centering
\caption{
  \label{tab:fluid-particle-val} Simulation parameters and setup for test cases to validate against experiments results by Ten Cate et al.~\cite{Ten_Cate2002-dg} and Mordant and Pinton~\cite{Mordant2000-lz}. Boundary conditions are periodic (p) or no-slip (ns).}
\begin{tabular}{lcccc}
\hline \hline
$Re_{t,\infty}$  & 12 & 32 & 41 & 360 \\\hline
Particle diameter $d_p$ (m)  & 0.015 & 0.015 & 1/6 & 1/6 \\
Density ratio $\rho_p/\rho_f$    & 1.16 & 1.16 & 2.56 & 2.56 \\
Fluid kinematic viscosity $\nu_f$ (10$^{-4}$~m$^2$/s)       & $1.17$ & $6.04$ & $54.2$ & $10.4$\\  \hline
Domain size (m) & $0.1 \times 0.1 \times 0.2$ &  $0.1 \times 0.1 \times 0.2$ & $1.25 \times 1.25 \times 10$ & $1.25 \times 1.25 \times 10$ \\
Grid resolution $d_p/h$ & 14, 19, 29 & 14, 19, 29 & 10, 20, 30, 43 & 10, 20, 30, 43 \\
Particle initial vertical position $z_0$ (m) & 0.13 & 0.13 & 9.5 & 9.5 \\
Boundary conditions & p $\times$ p $\times$ ns & p $\times$ p $\times$ ns & p $\times$ p $\times$ p & p $\times$ p $\times$ p \\
Outer forcing loops $n_f$ & 2 & 2 & 2 & 2 \\
Delta function & three-point & three-point & three-point & three-point \\
Pressure scheme ($\theta$) & Correction & Correction & Correction & Correction\\ \hline 
Reference & Ten Cate et al.~\cite{Ten_Cate2002-dg} & Ten Cate et al.~\cite{Ten_Cate2002-dg} & Mordant et al.~\cite{Mordant2000-lz} & Mordant et al.~\cite{Mordant2000-lz} \\ \hline \hline
\end{tabular}
\end{table}
\end{landscape}

\begin{figure}
\centerline{
 {\includegraphics[width=\textwidth]{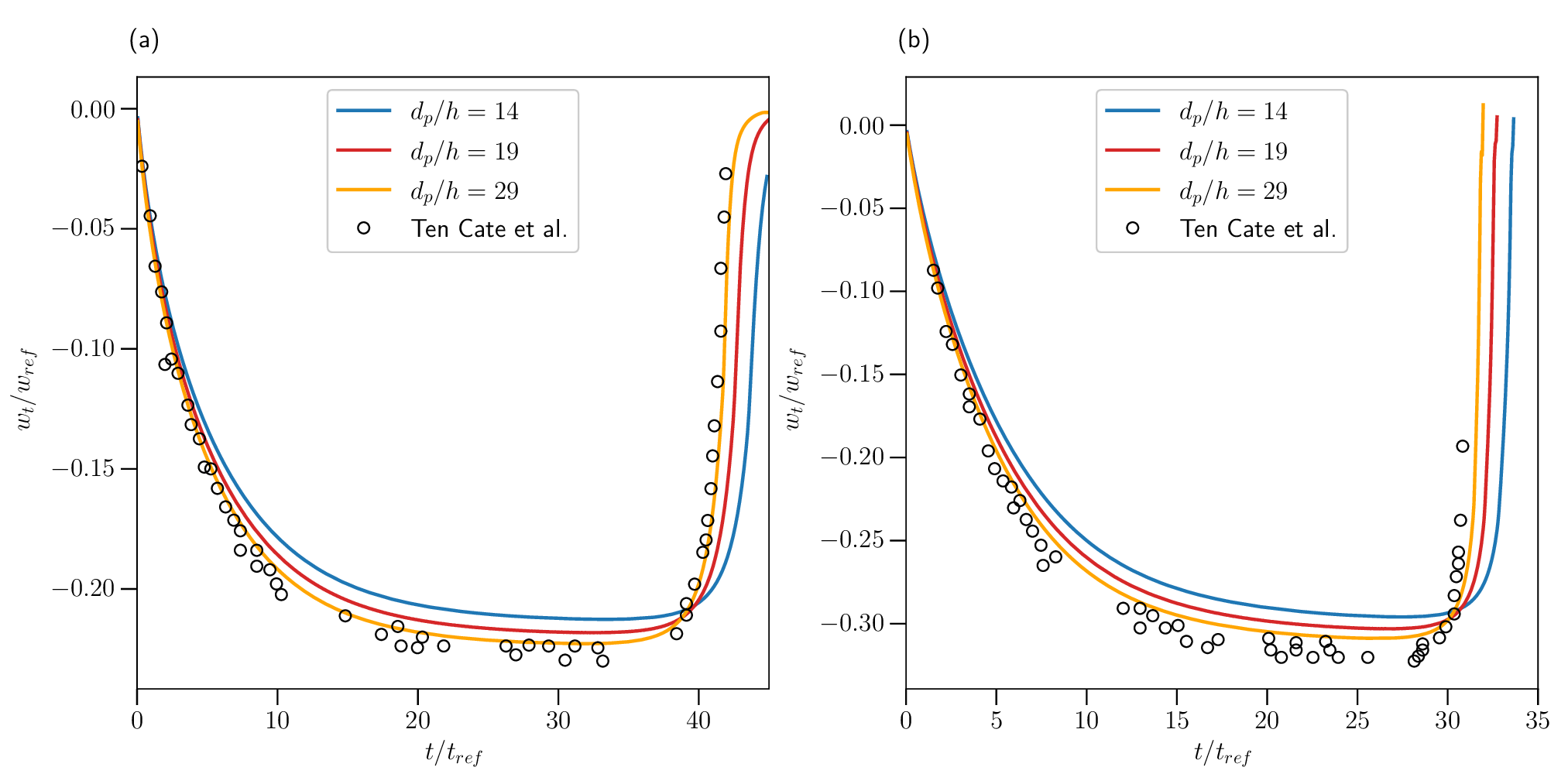}}}
\caption{Time series of the simulated settling velocity $w_{t}$ of a
 single particle settling and coming to rest on a bottom wall with $Re_{t,\infty}=12$ (a) and $32$ (b) and
 different grid resolutions $h$ used to resolve the particle diameter
 $d_p$, compared to published results.}
\label{fig:ten-cate}
\end{figure}

\begin{figure}
\centerline{
 {\includegraphics[width=\textwidth]{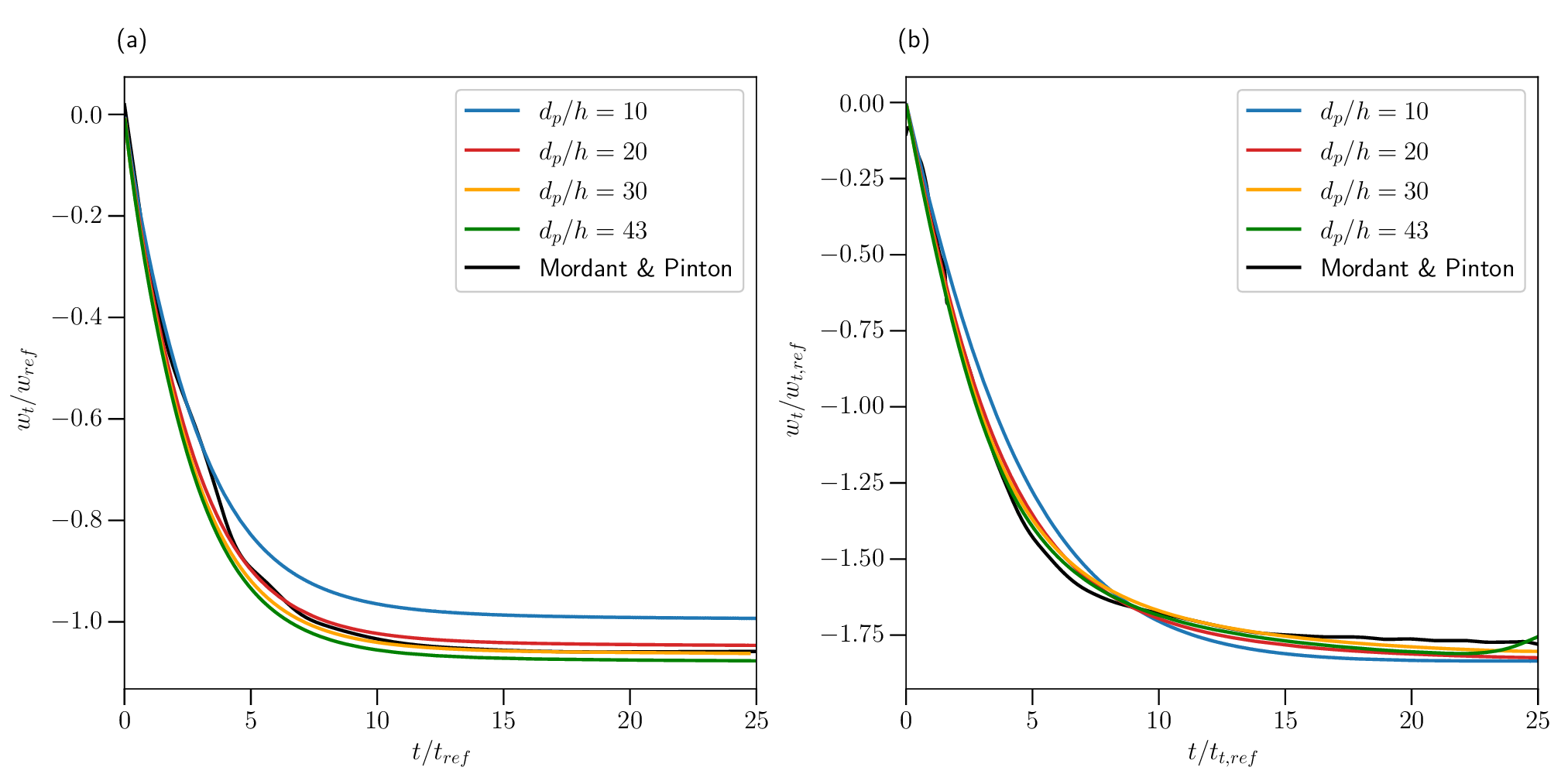}}}

\caption{Time series of the simulated settling velocity $w_{t}$ of a
 single particle in an approximately unbounded (periodic) domain with $Re_{t,\infty}=41$ (a) and $360$ (b) and
 different grid resolutions $h$ used to resolve the particle diameter
 $d_p$, compared to published results.}
\label{fig:mordant}
\end{figure}

Figures~\ref{fig:ten-cate} and~\ref{fig:mordant} show the settling velocity, $w_t$, of a single particle settling against a wall and in an approximately unbounded (periodic) domain, respectively, demonstrating the effect of grid resolution $d_p/h$. As the grid is refined and $d_p/h$ increases, the settling velocity $w_{t}$ converges monotonically to the experimental values. In Figure~\ref{fig:mordant}(b), the simulated settling velocity appears to exceed the experimental values, particularly for the higher $Re_{t,\infty}$ case. The experimental results by Mordant and Pinton~\cite{Mordant2000-lz} are ensemble averages of many different instances to average out small discrepancies in particle sizes and unsteadiness in the particle motion, which leads to a small discrepancy between experimental and simulation results and parameters. Similar discrepancies were also found by various authors~\cite{Uhlmann2005-hf,Biegert2018-rh,Breugem2012-rk}. Overall, based on these results, a grid resolution of $d_p/h=20$ achieves a good balance between computational cost and accuracy. Therefore, in what follows, we will use $d_p/h = 20$ as the default resolution.

\begin{figure}
\centerline{
 {\includegraphics[width=\textwidth]{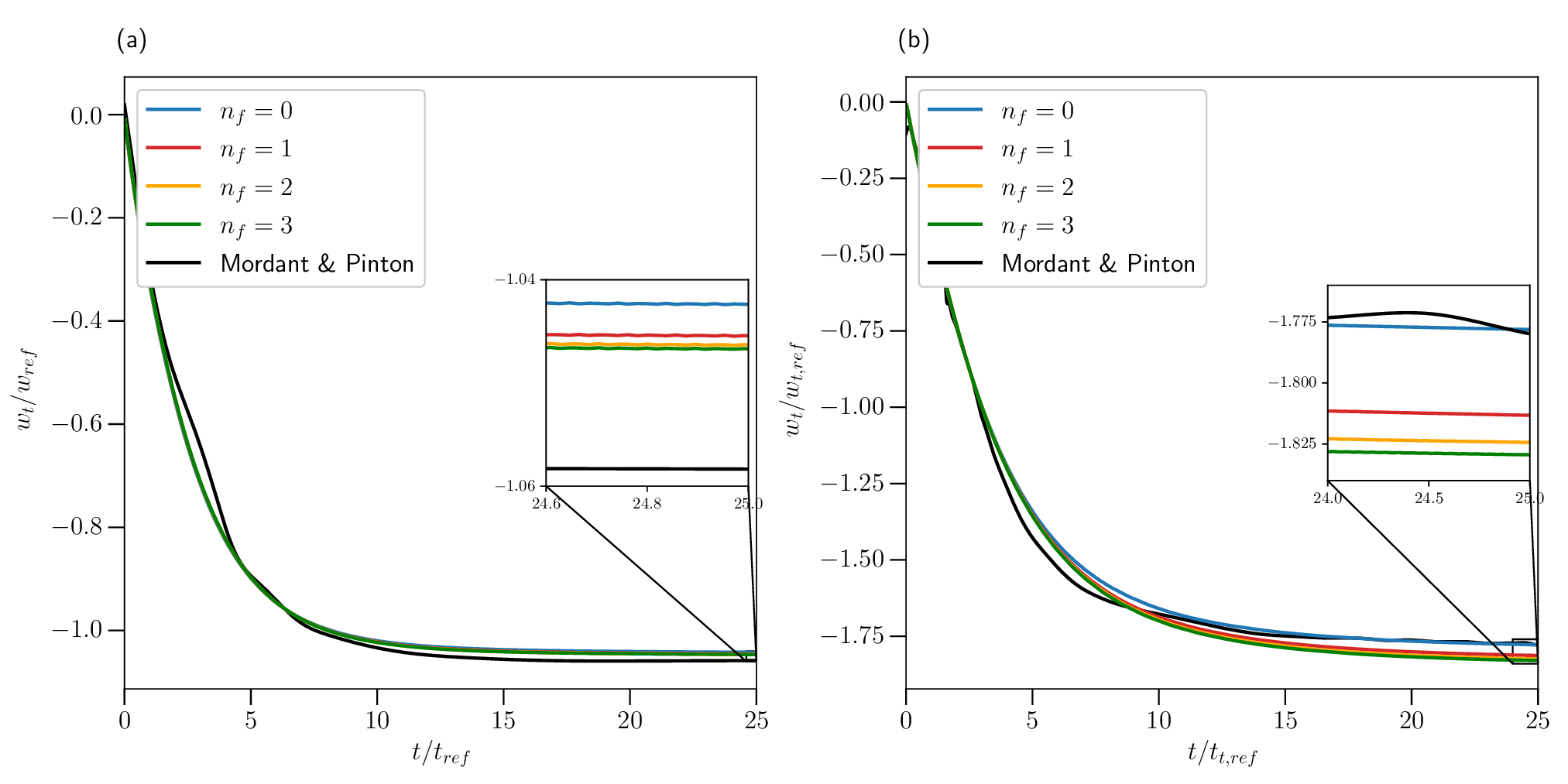}}}
\caption{Time series of the simulated settling velocity $w_{t}$ of a
 single particle with $Re_{t,\infty}=41$ (a) and $360$ (b) and
 different outer forcing loops $n_f$ to enforce the no-slip condition on the particle surface, compared to published results.}
\label{fig:nf-effect-mordant}
\end{figure}

In Section~\ref{sec:outer-f}, the number of outer forcing loops $n_f$ was introduced as a tuning parameter. To understand the effect of $n_f$ on the accuracy of fluid-particle interactions, simulations with different $n_f$ are conducted. Table~\ref{tab:fluid-particle-nf} summarizes the simulation parameters and Figure~\ref{fig:nf-effect-mordant} shows the effects of $n_f$ on the settling velocity $w_t$ in an approximately unbounded (periodic) domain. For the low Reynolds number case, $w_t$ converges towards the experimental results as $n_f$ increases (Figure~\ref{fig:nf-effect-mordant}(a)). Similar trends have been also observed by Kempe and Fr\"{o}hlich~\cite{Kempe2012-lp} and Biegert~\cite{Biegert2018-rh}. When $Re_{t,\infty} = 360$, the improvement from $n_f = 0$ to $n_f = 1$ is significant although further increasing $n_f$ does not demonstrate significant improvement (Figure~\ref{fig:nf-effect-mordant}(b)). The agreement between $n=0$ and experimental results in Figure~\ref{fig:nf-effect-mordant}(b) is likely to be coincidence since unsteadiness is introduced for higher Reynolds number. The experimental result for $Re_{t,\infty}=360$ is obtained by averaging numerous repetitions. Based on these simulations, $n_f = 1$ is sufficient to accurately simulate fluid-particle interactions. However, the number of outer forcing loops will also affect the accuracy of the collision models. Combined with studies conducted in Section~\ref{sec:par-par-interaction-val}, $n_f=2$ is required for accurate collision models while $n_f \ge 1$ is needed to obtain accurate fluid-particle interactions. Therefore, $n_f=2$ is appropriate to accurately simulate both fluid-particle and particle-particle interactions. 

\begin{table}[ht!]
\centering
\caption{
  \label{tab:fluid-particle-nf} Simulation setup to study the effect of the number of outer forcing loops $n_f$, pressure scheme, and Dirac delta function on simulations of particle settling. Boundary conditions are periodic in all directions.}
\begin{tabular}{ccc}
\hline \hline
$Re_{t,\infty}$  & 41 & 360 \\\hline
Particle diameter $d_p$ (m)  &  1/6 & 1/6 \\
Density ratio $\rho_p/\rho_f$    & 2.56 & 2.56 \\
Fluid kinematic viscosity $\nu_f$ (10$^{-4}$~m$^2$/s)    &  $54.2$ & $10.4$\\  \hline
Domain size (m $\times$ m $\times$ m)  & $1.25 \times 1.25 \times 10$ & $1.25 \times 1.25 \times 10$ \\
Grid resolution $d_p/h$ & 20 & 20 \\
Particle initial vertical position $z_0$ (m)  & 9.5 & 9.5 \\
Outer forcing loops $n_f$  & 0,1,2,3 & 0,1,2,3 \\
Delta function  & three- or four-point & three- or four-point \\
Pressure scheme ($\theta$)  & Correction, Projection & Correction, Projection\\ \hline \hline
\end{tabular}%
\end{table}

\begin{figure}
\centerline{
 {\includegraphics[width=\textwidth]{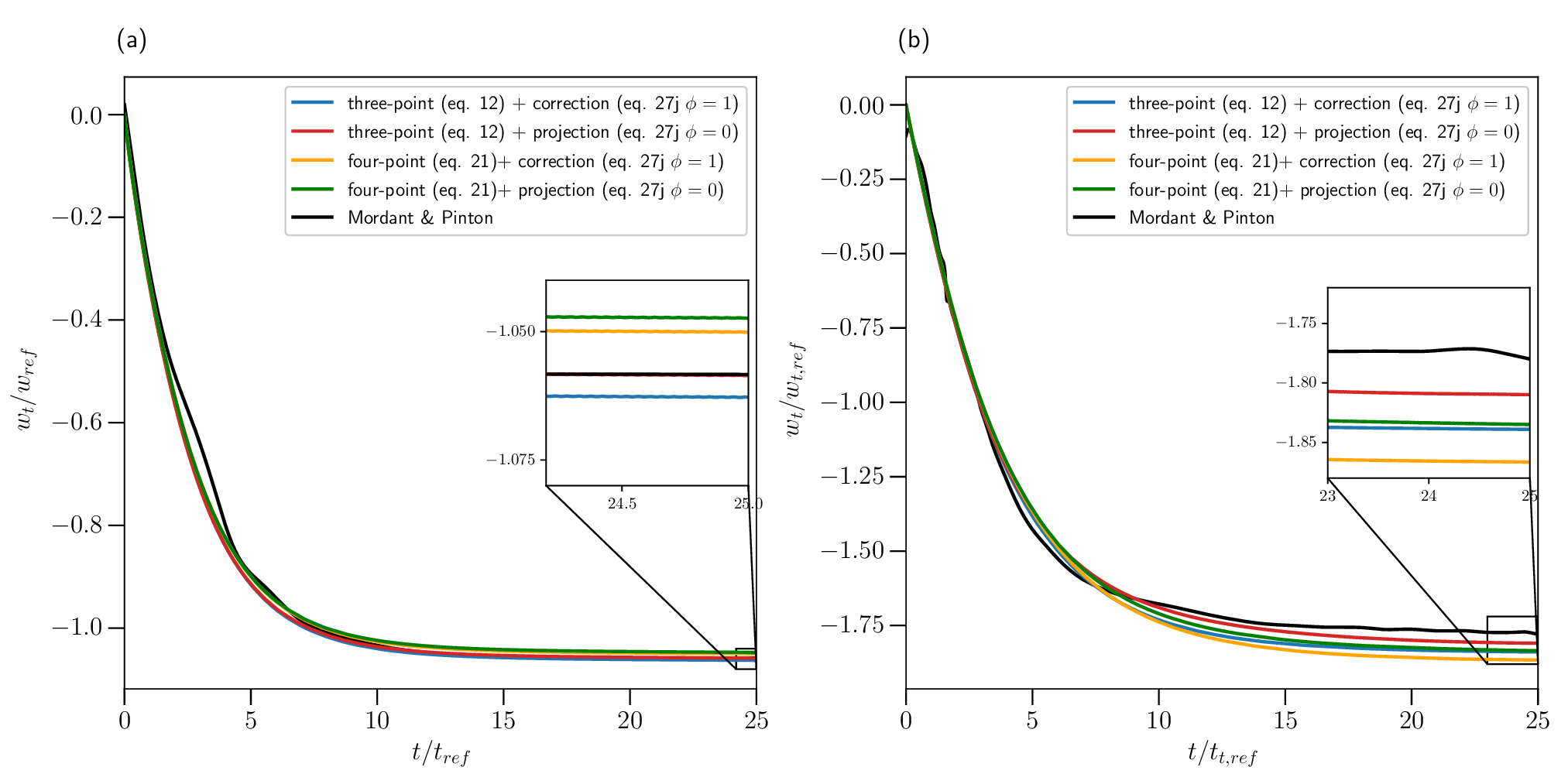}}}
\caption{Time series of the simulated settling velocity $w_t$ of a
 single particle with $Re_{t,\infty}=41$ (a) and $360$ (b) and
 different pressure schemes and delta functions, compared to published results.}
\label{fig:34pt-procor-Mordant}
\end{figure}

In addition to the grid resolution and $n_f$, choosing an appropriate interpolation kernel and pressure scheme has been shown to be critical, as discussed in Section~\ref{sec:modified-ibm}. Figure~\ref{fig:34pt-procor-Mordant} demonstrates the effects of 1) the pressure correction vs. projection scheme (equation~\ref{eq:fluid-solver}k) and 2) the three- vs. four-point regularized Dirac delta functions (equation~\ref{3pt-stencil} vs. equation~~\ref{eq:4-pt-stencil}). Although these parameters have been shown to have important effects by past authors (i.e. Armfield and Street~\cite{Armfield2002-zl} compared pressure correction and projection in staggered formulation of fractional step method without IBM and Uhlmann~\cite{Uhlmann2003-ke} tested three- and four-point Dirac delta functions in a staggered grid without collisions), thorough studies with the collocated IBM method have not been conducted. Our results show that a combination of the three-point delta function and pressure projection scheme produces results that most closely match the experiments for both low and high Reynolds numbers. In what follows, the three-point Dirac delta function and pressure projection scheme will be used.

\subsection{Particle-particle interactions}
\label{sec:par-par-interaction-val}
To validate and calibrate the normal collision model, simulation results are compared to the experiments of Gondret et al.~\cite{Gondret2002-ra}, in which a particle bounces off the wall of a tank. The experiments focus on the effect of the Stokes number on the maximum height of the particle after bouncing. The Stokes number is defined in equation~\ref{eq:stokes_num}. Table~\ref{tab:particle-particle-params} summarizes parameters and setup used in the simulations which were identical to Biegert et al.~\cite{Biegert2017-ku}, and we simulate cases with $St=27$ and $St=152$. Two tuning parameters to be calibrated are the number of outer forcing loops $n_f$ and minimum separation distance for the lubrication model $\zeta_{\min}$. The former governs the accuracy of the no-slip condition enforced on the particle surface while the latter determines the extent of deceleration due to lubrication forces. In the staggered direct forcing IBM~\cite{Biegert2017-ku}, $n_f = 1$ and $\zeta_{\min} = 3.0\times 10^{-3}r_p$ were chosen to reproduce experimental results by Gondret et al.~\cite{Gondret2002-ra} ($r_p = d_p/2$ is the particle radius).

To calibrate $n_f$ and $\zeta_{\min}$ for the collocated IBM in this paper, the particle is initially placed at a height 2 $d_p$ below the top wall and its velocity in the z-direction is prescribed according to Biegert et al.~\cite{Biegert2017-ku} as
\begin{eqnarray}
&& u \left(t \right) = -u_{imp}  \left(1-e^{-40t} \right).
\end{eqnarray}
This ensures a smooth acceleration to the desired impact velocity $u_{imp}$. The particle is then allowed to move freely once the distance to the bottom wall satisfies $\zeta_n < d_p/2$, whereupon the particle is subject to interaction with the fluid and wall.

\begin{table}[ht!]
\centering
\caption{\label{tab:particle-particle-params} Simulation parameters and setup to validate the collision models against the wall bounce experiments by Gondret et al.~\cite{Gondret2002-ra}. Boundary conditions are periodic (p) or no-slip (ns).}
\begin{tabular}{lcccc}
\hline \hline
Stokes number $St$  & 27 & 152  \\\hline
Particle diameter $d_p$(m)  & 0.006 & 0.003  \\
Density ratio $\rho_p/\rho_f$    & 8.083 & 8.342  \\
Kinematic viscosity $\nu_f$(10$^{-4}$~m$^2$/s)       & 1.036 & 0.107 \\
Restitution coefficient $e_{dry}$ & 0.97 & 0.97\\  
Impact velocity $u_{imp}$(m/s) &0.518 & 0.585 \\ \hline
Domain size (m $\times$ m $\times$ m)  & $0.08 \times 0.08 \times 0.16$ & $0.02 \times 0.02 \times 0.2$ \\
Grid resolution $d_p/h$ & 19.2 & 19.2 \\
Particle initial vertical position $z_0$ (m)  & 0.075 & 0.197 \\
Boundary conditions & p $\times$ p $\times$ ns & p $\times$ p $\times$ ns \\
Outer forcing loops $n_f$  & 0,1,2,3,5,10,20 & 0,1,2,3,5,10 \\
Dirac delta function  & three-point & three-point \\
Pressure scheme ($\theta$)  & Projection & Projection\\
Time step $\Delta{t}$ (s) & $2.5 \times 10^{-4}$ & $8.9 \times 10^{-5}$ \\ \hline \hline
\end{tabular}%
\end{table}

\begin{figure}
\centerline{
 {\includegraphics[width=\textwidth]{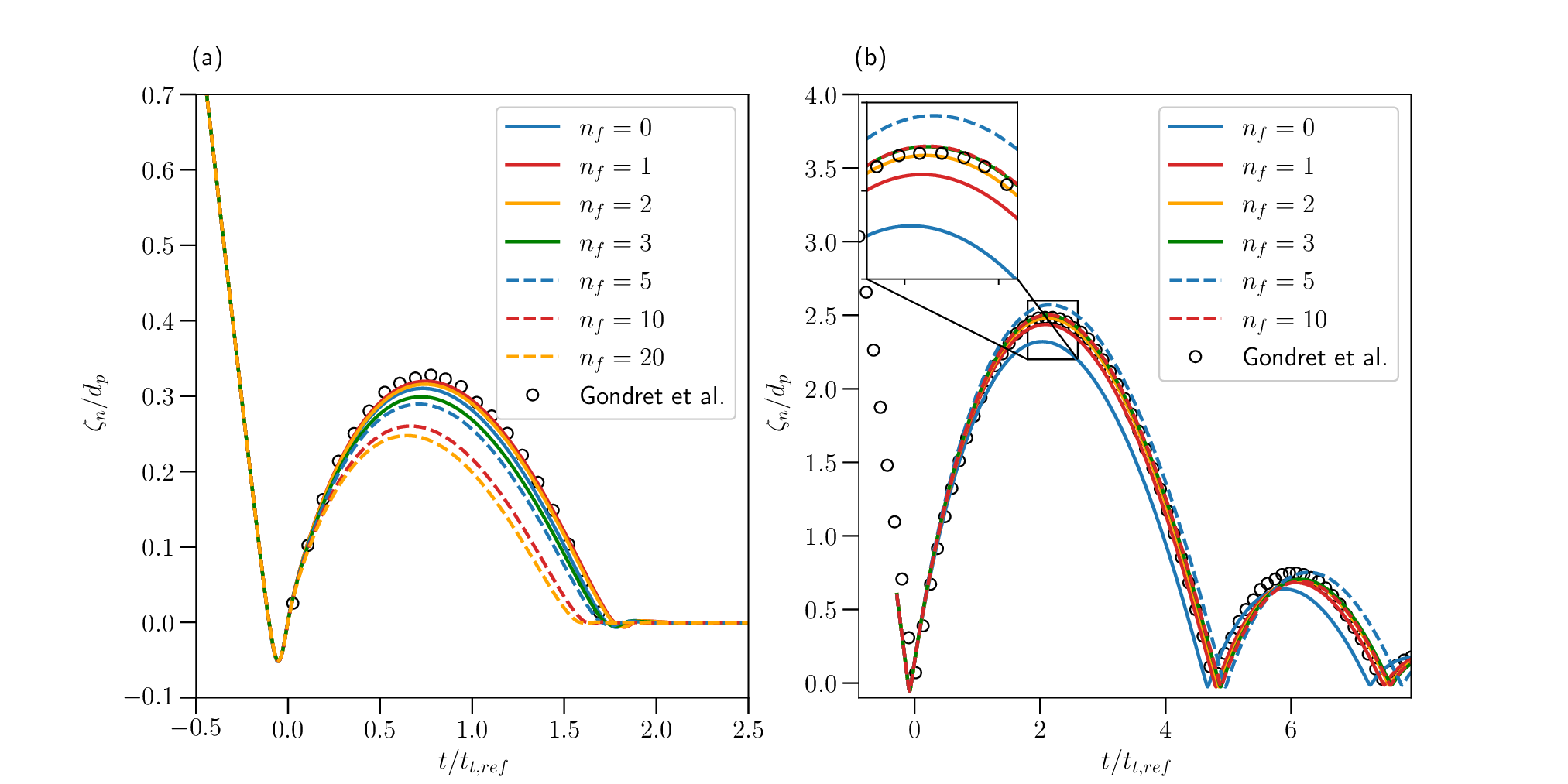}}}
\caption{Simulated height of the particle surface ($\zeta_n$) bouncing off of a wall with
 $St=27$ (a) and $152$ (b) with different number of outer forcing loops $n_f$ compared to published values. The minimum distance from the wall is 
 $\zeta_{\min} = 2.5 \times 10^{-3} r_p$.}
\label{fig:nf-effect-gondret}
\end{figure}
\begin{figure}
\centerline{
 {\includegraphics[width=\textwidth]{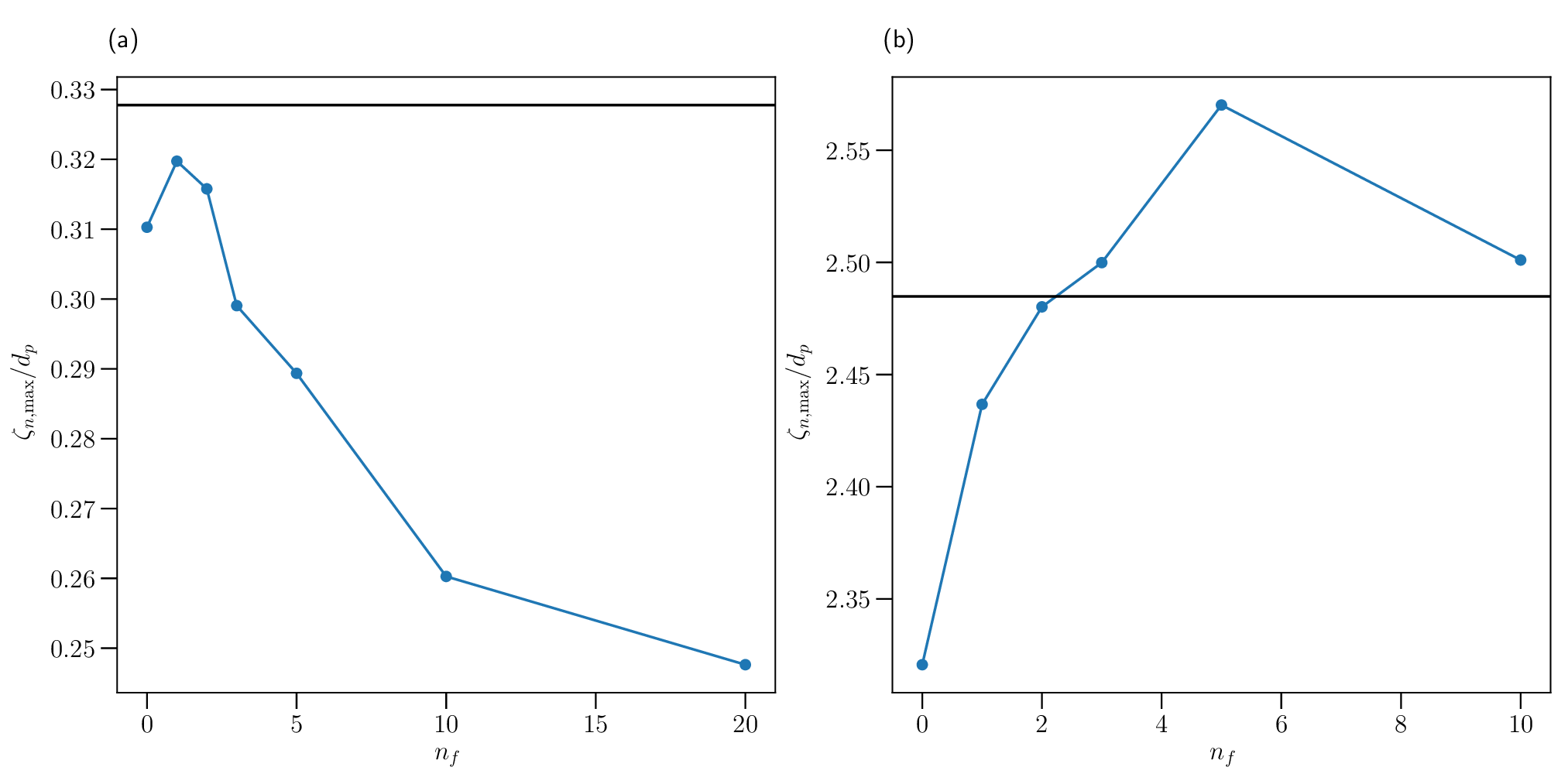}}}
\caption{Maximum rebouncing height of the center of a particle ($\zeta_{n,\max}$) with
 $St=27$ (a) and $152$ (b) as a function of number of outer forcing loops $n_f$. The black line represents the maximum rebouncing height from experiments by Gondret et al.~\cite{Gondret2002-ra}.}
\label{fig:zetamax-nf-gondret}
\end{figure}

To understand the effect of the number of outer forcing loops $n_f$ on the collision model, we conducted simulations with fixed $\zeta_{\min}$ and varied $n_f$ to reproduce the particle rebound height. Figure~\ref{fig:nf-effect-gondret} shows the effect of $n_f$ on the rebound height with $\zeta_{\min} = 2.5\times 10^{-3}r_p$. As shown in Figure~\ref{fig:zetamax-nf-gondret}, the maximum rebound height initially increases and then decreases with further increases in $n_f$, demonstrating the existence of a value of $n_f$ that achieves a maximum rebound height for a given $\zeta_{\min}$. 

\begin{figure}
\centerline{
 \includegraphics[width=0.45\textwidth]{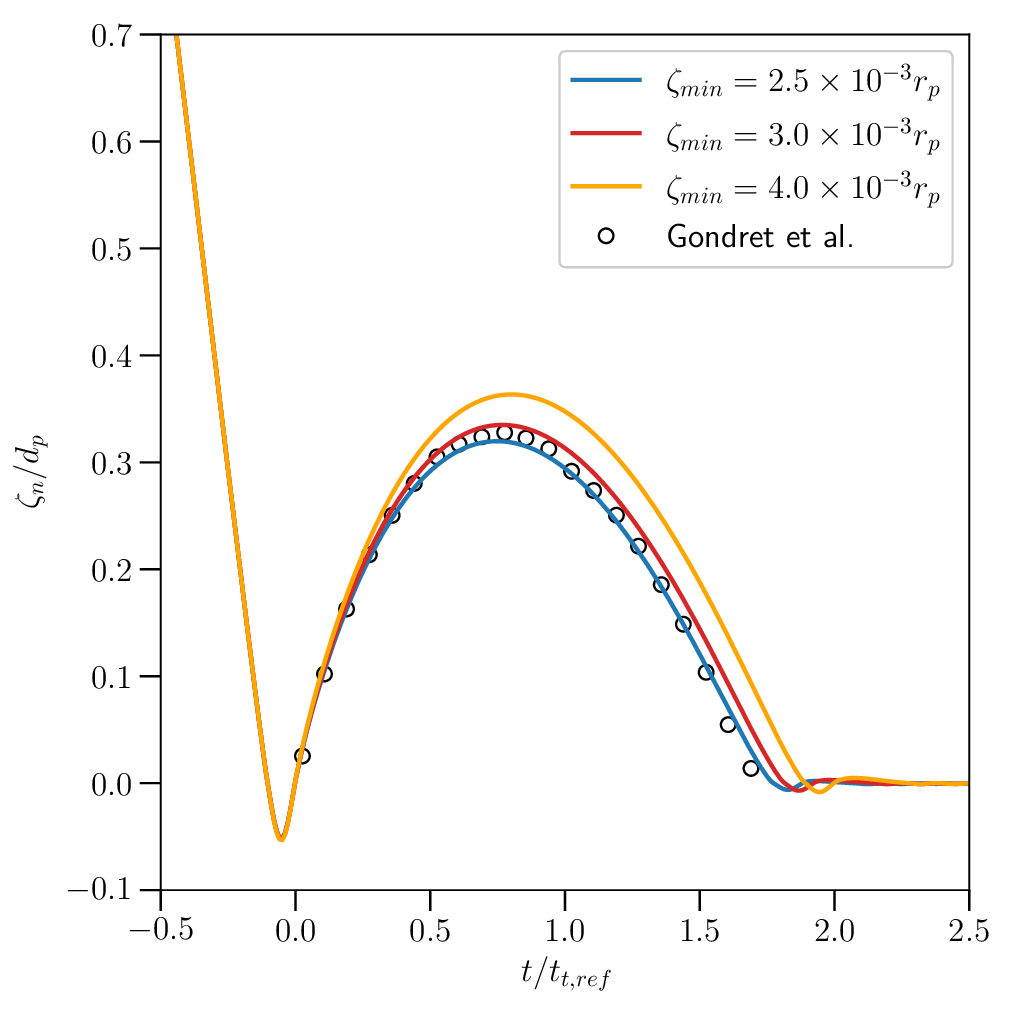}}
\caption{Simulated height of the particle surface ($\zeta_n$) bouncing off of a wall with
 $St=27$ with different $\zeta_{min}$ and $n_f = 1$ .}
\label{fig:zeteamin-nf1-effect-gondret}
\end{figure}

\begin{figure}
\centerline{
 {\includegraphics[width=\textwidth]{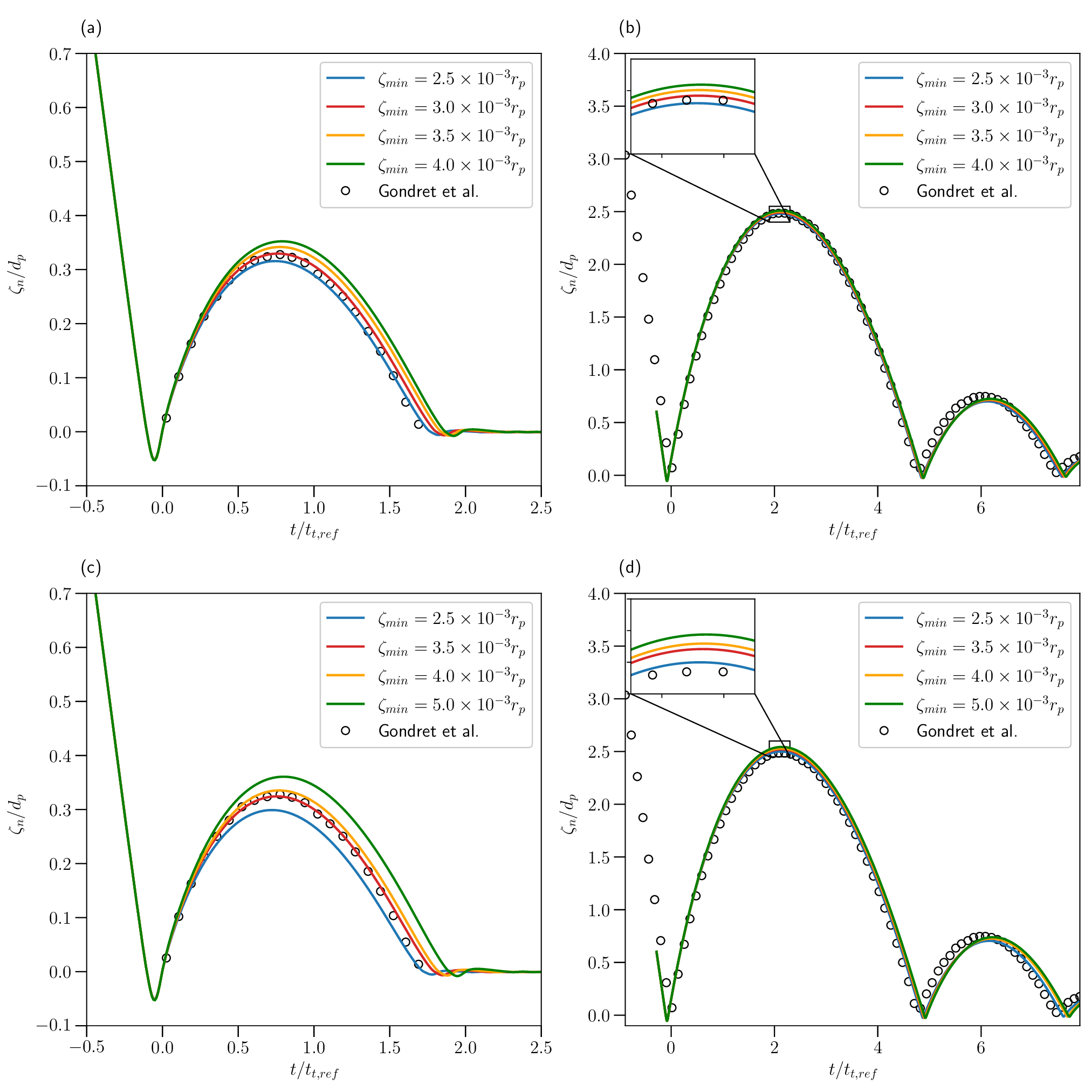}}
}
\caption{Simulated height of the particle surface ($\zeta_n$) bouncing off of a wall with
(a) $n_f=2$ and $St=27$, (b) $n_f=2$ and $St=152$, (c) $n_f=3$ and $St=27$ and (d) $n_f=3$ and $St=152$ with different $\zeta_{min}$.}
\label{fig:nf23-effect-gondret}
\end{figure}

Although there is a value of $n_f$ that maximizes the rebound height, the maximum value may not be the value that most closely matches the experiments because the value of $\zeta_{\min}$ is also important. To assess the effects of $\zeta_{\min}$, we simulated cases with varying $\zeta_{\min}$ and fixing $n_f = 1$ to study its effects on particle-particle interactions (The parameters are the same as those for case $St=27$ in Table~\ref{tab:particle-particle-params}). As shown in Figure~\ref{fig:zeteamin-nf1-effect-gondret}, $\zeta_{\min}$ has a significant effect on the rebound height because it dictates the lubrication force exerted on the particle. The smaller $\zeta_{\min}$, the greater the lubrication force experienced by the particle, hence the lower the particle rebound height. Unlike the effect of $n_f$, the rebound height decreases monotonically with decreasing $\zeta_{\min}$. The best values of $n_f$ and $\zeta_{\min}$ are obtained by conducting simulations with different $\zeta_{\min}$ for $n_f = 1,2$ and $3$ and choosing values that best match the experiments for both $St=27$ and $St=152$. The computed errors for all cases simulated are summarized in Table~\ref{tab:zetamax_nf_error}. For $n_f = 2$ and $n_f = 3$ as shown in Figure~\ref{fig:nf23-effect-gondret}, the values of $\zeta_{\min}$ to obtain the most accurate rebound heights are $3.0\times 10^{-3}r_p$ and $3.5\times 10^{-3}r_p$, respectively. Since $n_f = 2$ is already satisfactory for fluid-particle interaction, $n_f = 2$ and $\zeta_{\min} = 3.0\times 10^{-3}r_p$ are chosen for both computational cost and accuracy.

\begin{table}[ht!]
\centering
\caption{\label{tab:zetamax_nf_error} Summary of error $\abs{\zeta_{n,\max,sim} - \zeta_{n,\max,exp}}$ between simulations and experiments by Gondret et al.~\cite{Gondret2002-ra} as a function of $\zeta_{\min}$ and $n_f$.}
\begin{tabular}{cccccc}
\hline \hline
$\zeta_{min}$   & $n_f=1$ & \multicolumn{2}{c}{$n_f=2$} & \multicolumn{2}{c}{$n_f=3$} \\ \hline
 & $St=27$ &  $St=27$ & $St =152$ & $St=27$ & $St =152$ \\ \hline
$2.5\times 10^{-3}r_p$  & 0.008 & 0.012 & 0.005 & 0.029 & 0.015 \\
$3.0\times 10^{-3}r_p$  & 0.007 & 0.002 & 0.007 & -     & -     \\
$3.5\times 10^{-3}r_p$  & -     & 0.014 & 0.016 & 0.003 & 0.036 \\
$4.0\times 10^{-3}r_p$  & 0.036 & 0.025 & 0.025 & 0.008 & 0.045 \\  
$5.0\times 10^{-3}r_p$  & -     & -     &  -    & 0.033 & 0.059 \\ \hline
\end{tabular}%
\end{table}

\subsection{Strong and weak computational scaling}
\label{sec:scaling}
The code is parallelized with MPI (openmpi-4.0) and we used the Hypre
libraries developed by Lawrence Livermore National
Laboratories~\cite{Falgout2006-bt} to invert the large linear systems
associated with the fluid pressure and viscous terms. Lagrangian
particle information is transferred between processors using an MPI
struct. We used Stampede2 KNL (TACC at University of Texas, Austin) with Intel Xeon Phi 7250 processors (1.4 GHz) from XSEDE~\cite{Towns2014-xm} to obtain the scaling results outlined below. 

The scaling test cases were conducted with a fluidized-bed reactor that is periodic in the \emph{x}- and \emph{y}-directions with inflow and outflow conditions at the top and bottom boundaries (in the \emph{z}-direction; see Section~\ref{sec:mono-sim} for details). In all simulations, the particle diameter is $d_p=$2~mm, the upward flow velocity is 0.05~m~s$^{-1}$ and the domain size in the \emph{x}, \emph{y} and \emph{z} directions is $15d_p\times 15 d_p\times 45 d_p$. Ten time steps were computed with a time-step size of $10^{-4}$~s, and results are reported as the average wall-clock time per time step. 

Strong scaling of the code is assessed by fixing the number of particles and number of grid points, and varying the total number of MPI tasks. To demonstrate strong scaling behavior of our code, we conducted three-dimensional simulations with 2000 spherical particles. The particles were uniformly distributed in the domain or closely packed near the bed to provide worst- and best-case scaling scenarios. The total problem size was either $256\times 256 \times 768$ or $512 \times 512 \times 1536$ Eulerian grid points. 

\begin{figure}[ht!]
\centering
\includegraphics[width=0.5\textwidth]{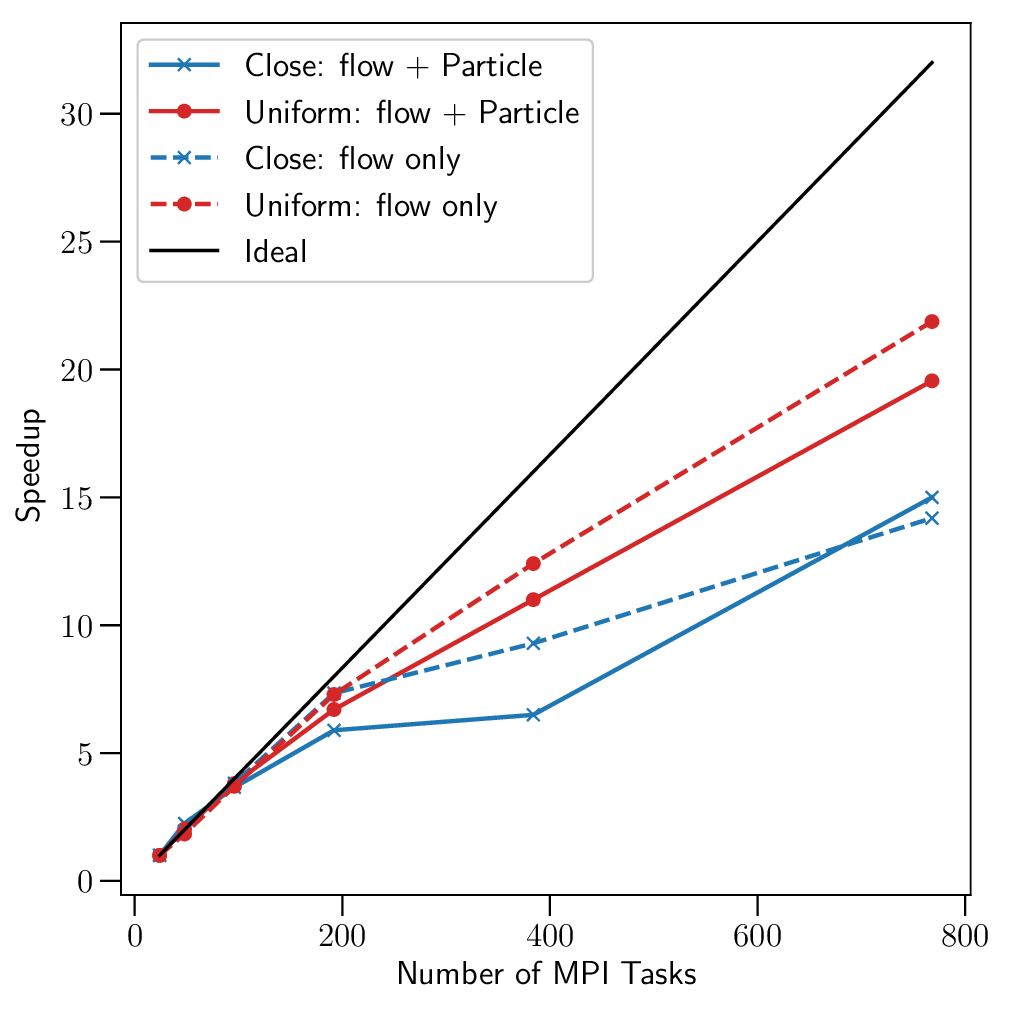}
\caption{Strong scaling with 2000 particles in a doubly-periodic cubic domain, showing how the particle solver is not as efficient as the flow solver, and the efficiency is better for the uniform distribution of particles owing to improved load balancing.}
\label{fig:strong_long}
\end{figure}

In order to compare the parallel performance of the components of the
code related to the flow and particle solvers, we define the total wallclock
time with $n$ MPI tasks as the sum of the time needed for the flow and
particle calculations as $t_n = t_{flow,n} + t_{particles,n}$. The
speedup related to calculation of the flow+particles is then given by
$S_{flow+particles,n}=t_{24}/t_n$, while the speedup related to the flow solver
only is $S_{flow,n}=t_{flow,24}/t_{flow,n}$, where $n \ge 24$ is chosen due to memory limitation. As shown in
Figure~\ref{fig:strong_long}, since there are substantially fewer
particles than flow grid cells, the parallel efficiency of the
particle solver is smaller owing to the relatively fine-grained parallelism for the particle
calculations. This effect is pronounced when the simulations are initialized
with a closely packed bed, in which case load balancing is 
less efficient because the particle workload is disproportionately assigned to
processors containing the closely packed particles.

In weak scaling, the typical approach is to keep the number of grid
points on each processor constant. This is achieved by varying the
number of processors and total number of grid points
simultaneously. In our method, weak scaling is demonstrated
by keeping both the number of particles and grid points on each subdomain constant. However, a
challenge with the weak scaling is that the computational cost of
the particle solver increases with grid refinement related to the flow solver
because the particle surfaces are resolved with more Lagrangian markers that
must be coincident with the Eulerian grid cells. Therefore, weak scaling disproportionally
adds more work to the particle solver when refining the Eulerian grid. To ensure
that the workload for each processor related to the particle solver is
fair, we reduce the particle diameter in proportion to the grid
spacing when refining the grid.
For example, refining the grid from $64\times64\times192$ to
$128\times128\times384$ would reduce the particle diameter $d_p$ from 2~mm to
1~mm. This ensures that the number of Lagrangian markers needed to simulate
the fluid-particle interactions remains constant on each processor.

\begin{figure}[ht!]
\centering
\includegraphics[width=\textwidth]{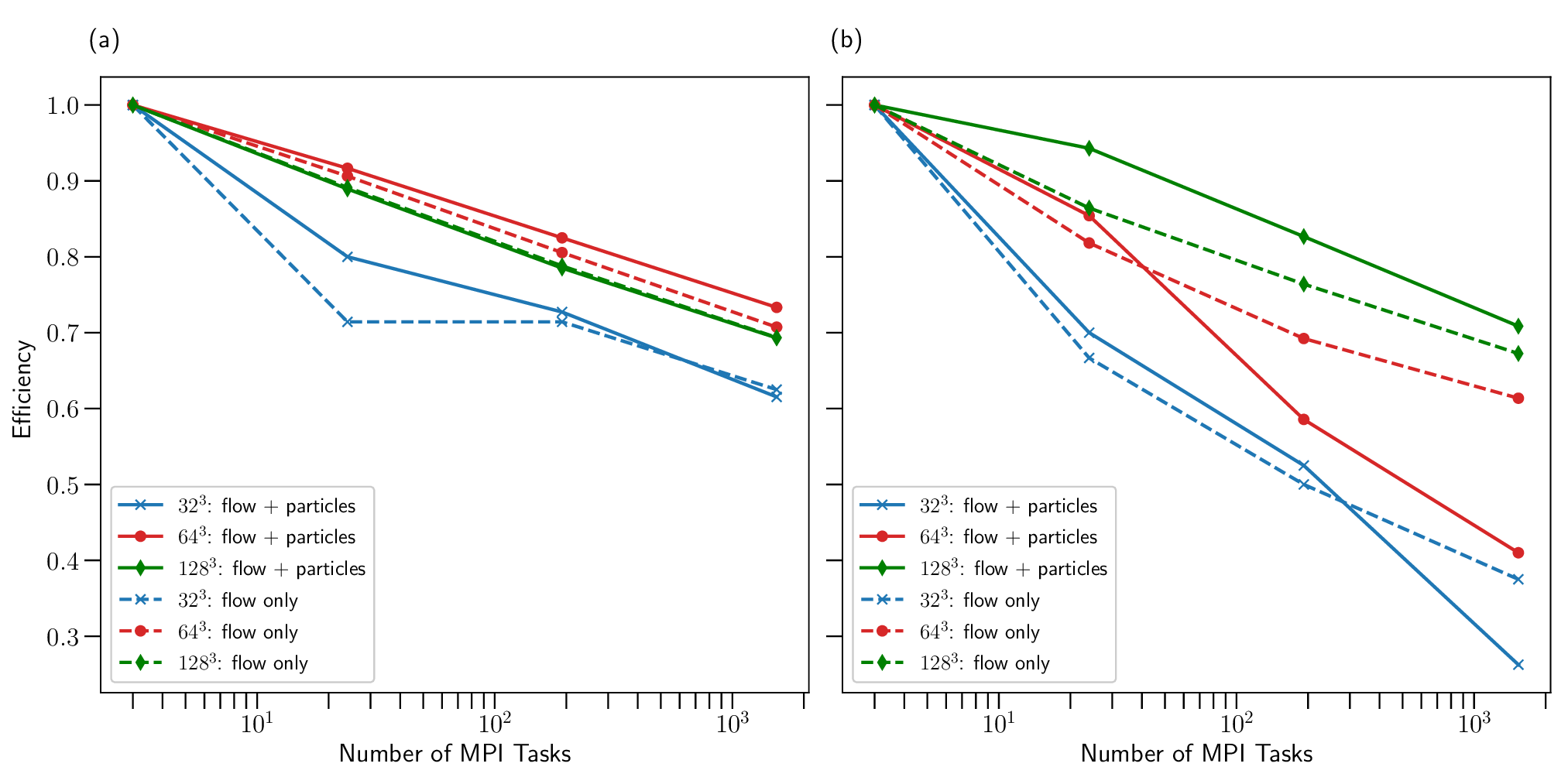}
\caption{Weak scaling for the flow solver and for the flow+particle solver with different grid sizes on each processor for a doubly-periodic cubic domain. The scaling for actual simulations is expected to be somewhere between the best-case (a) and worst-case (b) particle scenarios.}
\label{fig:weak_scaling}
\end{figure}

Weak scaling is demonstrated for different grid sizes on each processor and
with two different initial locations
of the 125 particles per processor:
(a) all particles are clustered in the center of each subdomain
so that no information is exchanged, and (b) most particles are  distributed along the boundaries of the 
processor so that particle information on the boundaries must be exchanged.
As shown in Figure~\ref{fig:weak_scaling}, scaling is best when using $64^3$
grid points per MPI task, although the performance is not as good
for case (b). However, typical cases are expected to have 
scaling behavior that is somewhere between cases (a) and (b).

Overall, the weak scaling results indicate that a simulation with 400 million grid cells is expected to
take roughly 7~s per time step using ~1500 MPI tasks (assuming behavior that is half-way between
cases (a) and (b)). Extrapolation indicates that a simulation with 1~billion grid cells should take 9~s
per time step with 3000 MPI tasks, and a simulation with 5~billion grid cells should take roughly 20~s per time step

\subsection{Monodispersed and bidispersed particle fluidization}
\label{sec:mono-sim}
To demonstrate the capability of the method to simulate fluidized bed reactors, we conducted simulations showing that we can match the porosity ($1-\phi$ where $\phi$ is the volume fraction) predicted by experiments. In a fluidized bed, the porosity can be predicted with the power law relationship 
\begin{eqnarray}
\label{eq:pow-k}
&& u^* = \frac{u_0}{w_{t,ref}} = k  \left(1-\phi \right)^n,
\end{eqnarray}
where $u^*$ is the normalized velocity, $u_0$ is the superficial or upflow velocity, $w_{t,ref}$ is settling velocity of a single particle in the domain of interest, $k= 0.7-0.9$ is a constant prefactor~\cite{Yao2021-ex, Yin2007-eb, Willen2019-rm, Yao2021-rg} and $n$ is the power law exponent. The general consensus in the literature is that Equation~\ref{eq:pow-k} can predict the hindered settling velocity for a particle suspension or the porosity for fluidization. Researchers have established various relationships to relate $n$ to the terminal Reynolds number of a single particle  given by equation~\ref{eq:Retinf}~\cite{Garside1977-vp, Richardson1954-ay}. Richardson and Zaki~\cite{Richardson1954-ay} employ the stepwise function
\begin{eqnarray}
&& n = \label{eq:Zaki}
\begin{cases} 
4.65, & Re_{t,\infty} < 0.2\,,\\
4.35Re_{t,\infty}^{-0.03}, & 0.2 \le Re_{t,\infty} < 1\,,\\
4.45Re_{t,\infty}^{-0.1}, & 1 \le Re_{t,\infty} < 500\,,\\
2.39, & Re_{t,\infty}\ge 500\,.
\end{cases}
\end{eqnarray}
Garside and Al-Dibouni~\cite{Garside1977-vp} improved the relationship with a continuous sigmoid function proposed to relate $n$ and the terminal Reynolds number of a single particle in the domain of interest, $Re_{t,ref} = w_{t,ref}d_p/\nu_f$ where $w_{t,ref}$ is the terminal velocity of a single particle in the domain of interest, as 
\begin{eqnarray}
&& \frac{5.1-n}{n-2.7} = 0.1 Re_{t,ref}^{0.9}\,,\label{eq:Garside}
\end{eqnarray}
which is 20\% more accurate than Equation~\ref{eq:Zaki}~\cite{Yin2007-eb}. 

\begin{figure}
\centering
\includegraphics[height = 0.5\textheight, keepaspectratio]{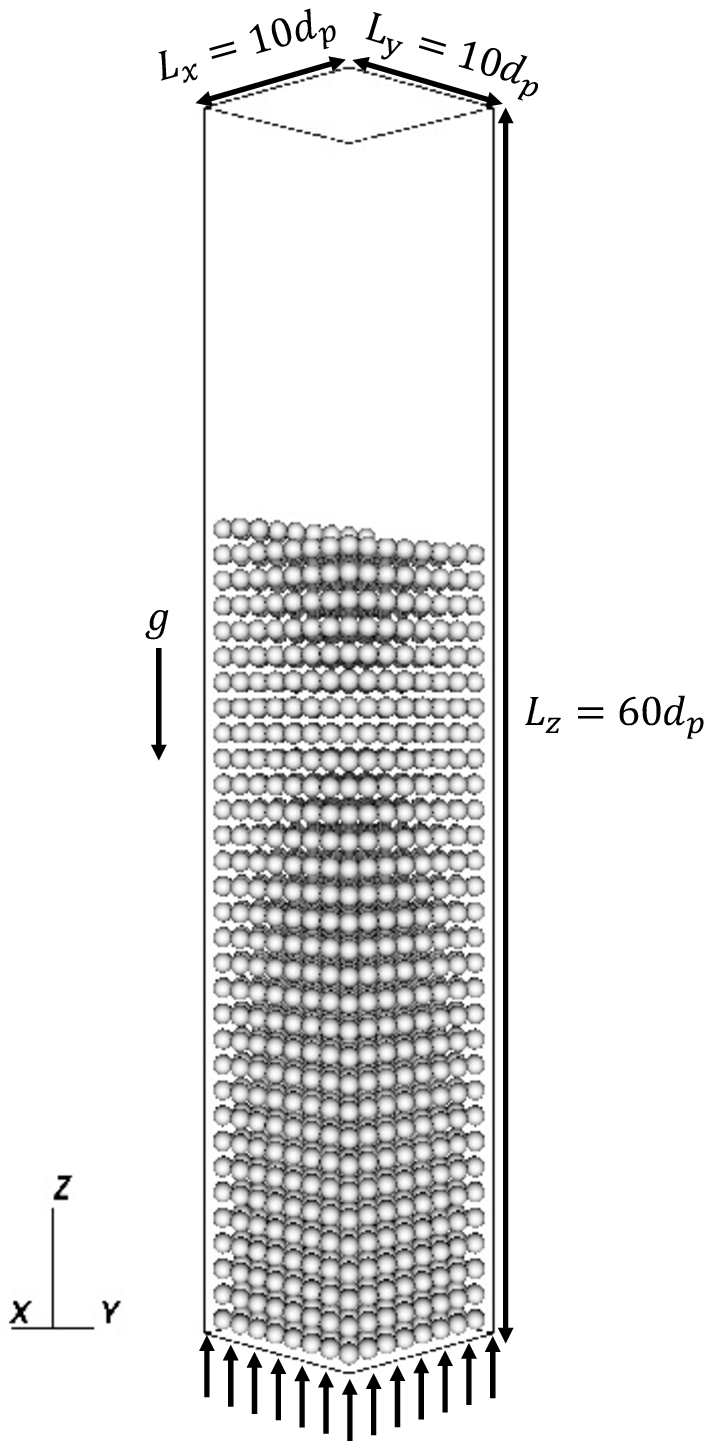}%
\caption{\label{sch_diag}Schematic of the three-dimensional
 computational domain, showing the initial particle locations making up the fluidized bed.}
\end{figure}

To verify that our method can reproduce the power law~\ref{eq:pow-k}, three-dimensional simulations are conducted with $N_p=2000$ particles in the
reactor channel shown in Figure~\ref{sch_diag}. The particles have a uniform diameter $d_p=0.002$~m and density $\rho_p=1300$~kg~m$^{-3}$,
and the fluid has a kinematic viscosity $\nu_f=1.0037 \times 10^{-6}$~m$^2$~s$^{-1}$
and density $\rho_f=998.21$~kg~m$^{-3}$ ($\rho_p/\rho_f = 1.3$). The grid spacing is uniform
in the $x$, $y$ and $z$ directions and given by $\Delta x= \Delta{y} = \Delta z=h=d_p/25.6$, which is sufficient to resolve the flow-particle interactions as demonstrated
in Section~\ref{sec:vals}. The square channel dimensions are given by $L_x = L_y = 10 d_p$ and its length is $L_z = 60 d_p$, giving
a three-dimensional grid with 256$\times$256$\times$1536 grid points.
The time-step size is $\Delta{t}=1.5 \times 10^{-4}$~s, resulting in a maximum Courant number of $0.4$ for the cases with the highest upflow velocities. In all simulations, 384 processors were used ensuring a $64^3$ computational domain for each processor. Each time step requires 10 s wall clock which is consistent with the scaling results presented in Section~\ref{sec:scaling}.

Simulations are initialized with a uniform distribution
of close-packed particles with a spacing of $1 d_p$ at the bottom of the domain and the flow is impulsively started from rest. The upflow velocity leads to expansion of the bed and random motion of the particles until statistical equilibrium is reached, at which time the dynamics are independent of
the initial particle distribution. Simulations are run for a total of 100 $\tau_T$, where $\tau_T = d_p/u_0$ is the particle turnover time. As it takes roughly 30 $\tau_T$ to reach statistical equilibrium, results are time averaged over the last 70 $\tau_T$.

The average upflow velocity at the inlet, $u_0$, is varied to investigate Reynolds number effects. A total of six simulations were conducted with $0.01\le u_0\le 0.35$~m~s$^{-1}$, giving $20 \le Re_p \le 70$ where $Re_p = u_0 d_p /\nu_f$. For all cases, the pressure is specified at the top boundary as $p=0$,
while at the bottom boundary the inflow velocity is specified and all side wall boundary conditions are periodic. Due to the absence of walls which would produce a Poiseuille velocity profile at the inlet, the inflow velocity is uniform and given by $u_0$. For collision models, the minimum separation distance in the lubrication model is set to $\zeta_{\min} = 3.0\times10^{-3}r_p$ and the dry restitution coefficient $e_{dry} = 0.97$. The parameters dictating rolling/sticking $\mu_s = 0.11$  and sliding $\mu_k = 0.8$ are used.

\begin{figure}
\centerline{
 \includegraphics[width=0.45\textwidth]{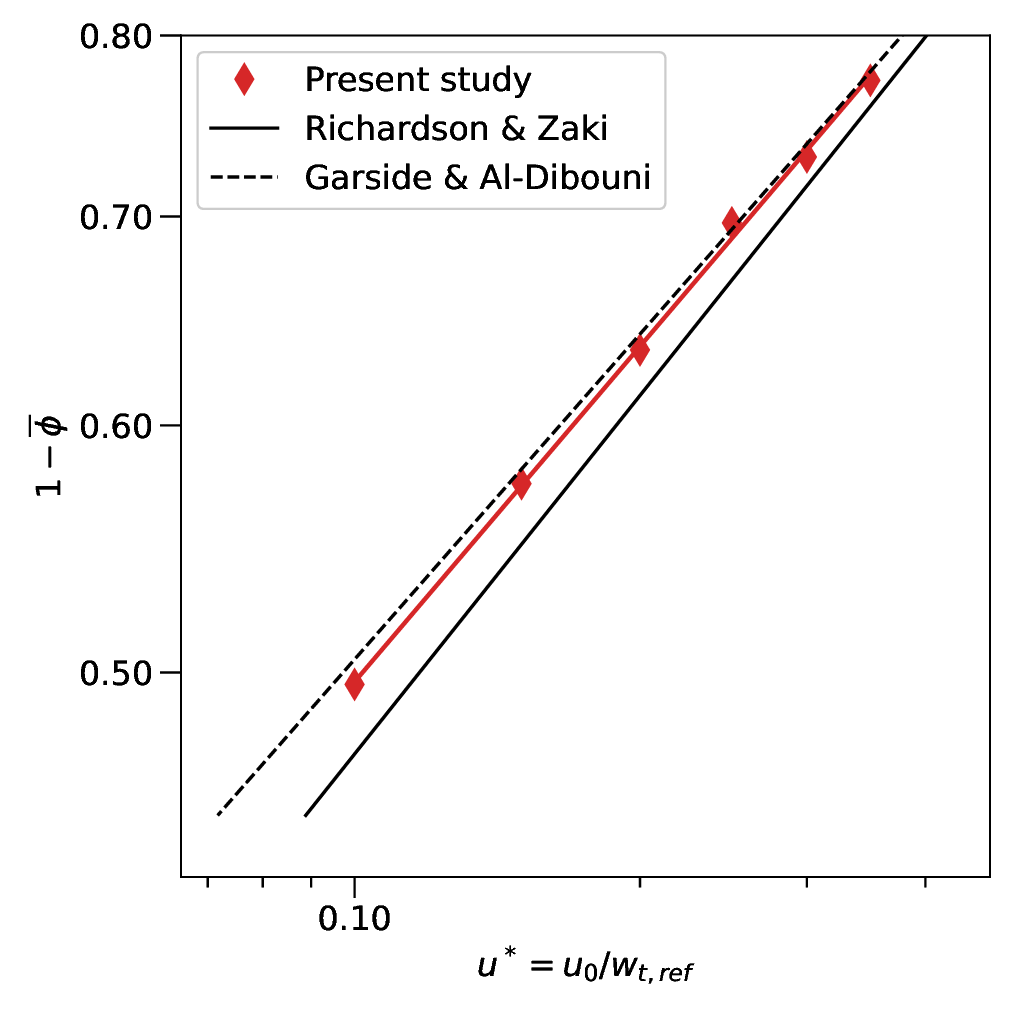}}
\caption{Time-averaged porosity $1-\overline{\phi}$ as a function of superficial velocity $u_0$ normalized by the settling velocity of a single particle in the domain of interest $w_{t,ref}$ for the simulated cases. The lines were constructed based on fitting to the power law equation~(\ref{eq:pow-k}).}
\label{fig:expansion-fit}
\end{figure}

Figure~\ref{fig:expansion-fit} demonstrates the relationship between the normalized velocity $u^*$ and the time-averaged porosity $1-\overline{\phi}$. A straight line on a log-log scale indicates the results follow a power law relationship. To assess the accuracy of the simulations, we regressed the power law (equation~\ref{eq:pow-k}) which gives $k=0.72$ and $n=2.82$. This is in close agreement with the values of $n = 2.62$ (equation~\ref{eq:Zaki}) and $n= 2.88$ (equation~\ref{eq:Garside}) and trends reported by other low-porosity particle-resolved simulations~\cite{Yin2007-eb,Willen2019-rm}.

\begin{figure}[ht!]
\centering
\includegraphics[width=0.5\textwidth]{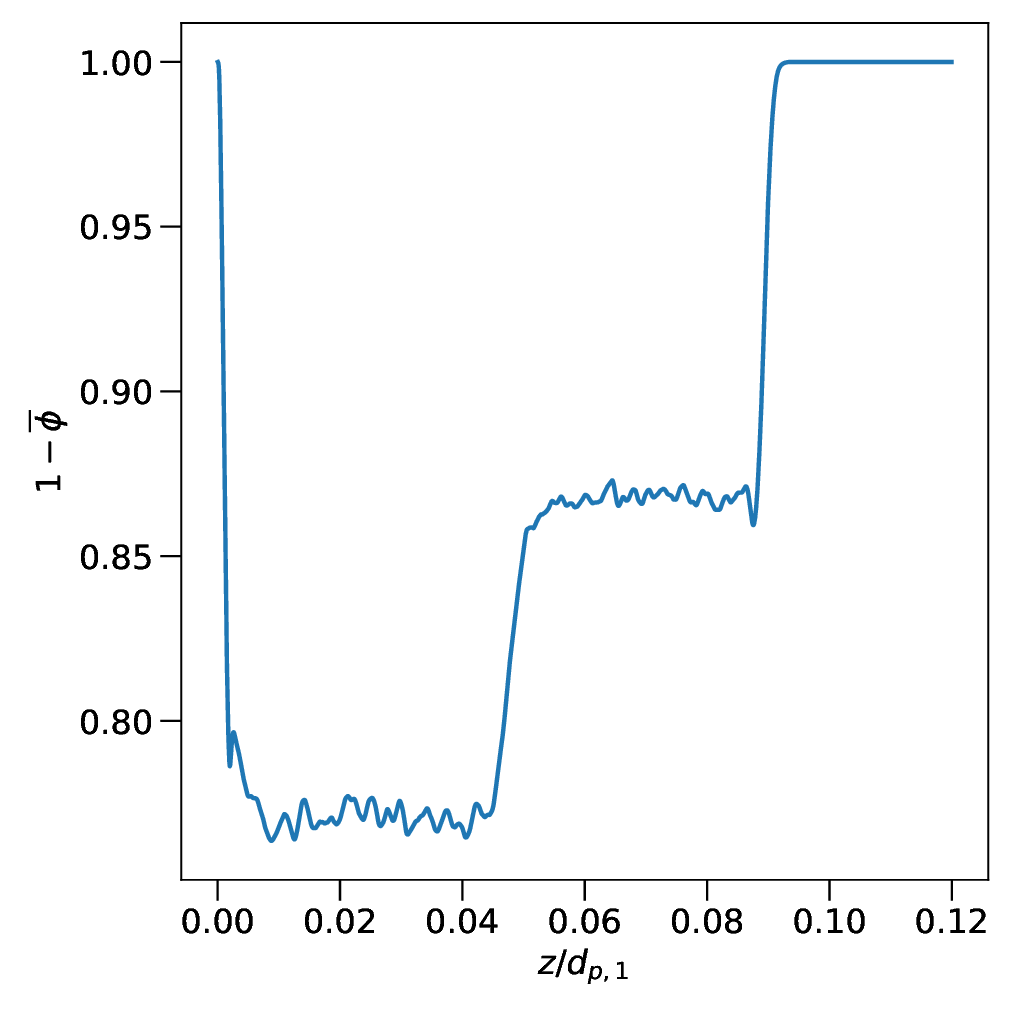}
\caption{Time-averaged porosity $1-\overline{\phi}$ as a function of vertical position z for an upflow velocity of 0.35~m~s$^{-1}$ in a bidispersed fluidized bed.}
\label{fig:bi-dis-phi}
\end{figure}

To demonstrate the capability to simulate polydispersed particles, we increased the simulation complexity from monodispersed to bidispersed fluidization. Here, we performed a three-dimensional simulation with $N_{p,total}=3376$ particles in the reactor. The particles have two uniform diameters $d_{p,1}= 0.002$~m and $d_{p,2}=d_{p,1}/1.4$ and density $\rho_p=1300$~kg~m$^{-3}$. The number of particles for $d_{p,1}$ and $d_{p,2}$ are $1000$ and $2376$, respectively. The fluid properties are identical to the monodispersed fluidization simulations. A uniform grid spacing with $h/d_{p,1}=25.6$ was used to ensure accuracy. The square channel dimensions were identical, giving a grid with 256$\times$256$\times$1536 points. The inflow velocity was 0.35~m~s$^{-1}$, giving $Re_p$ calculated based on $d_{p,1}$ as 70. The time-step size $\Delta{t}$ is calculated with both advection and diffusion Courant numbers that are defined as $C_{adv}=u_0\Delta{t}/h$ and $C_{diff}=\nu_f\Delta{t}/h^2$, respectively. We ensure that the maximum Courant number $C_{max} = max(C_{adv}, C_{diff}) = 0.25$. Figure~\ref{fig:bi-dis-phi} shows the time-avergaed porosity $1-\overline{\phi}$ as a function of $z$. The segregated bidispersed fluidized bed consists of three regions with the following properties, in order from the bottom to the top of the reactor: (1) monodispersed fluidization with particles of size $d_{p,1}$, (2) bidispersed fluidization consisting of both particle sizes, and (3) monodispersed fluidization with particles of size $d_{p,2}$. We computed the spatially-averaged porosity for the lower and upper regions as $1-\overline{\phi}=0.771$ and $0.868$, respectively which are approximately equivalent to the time-averaged porosity obtained for the equivalent monodispersed fluidization simulations ($1-\overline{\phi}=0.773$ and $0.869$). These results indicate that bidispersed fluidized beds essentially segregate into layers that behave like their monodispersed equivalents, as discussed extensively in a paper using the present method by Yao et al.~\cite{Yao2021-ky}.

\section{Conclusion} 
We develop a modified particle-resolved simulation framework that combines the collocated-grid Immersed Boundary Method and colision models for polydisperse particles. An improved Immersed Boundary Method~\cite{Kempe2012-lp} is coupled with the fractional step method~\cite{Zang1994-ck} to solve the Navier Stokes equations. The method is shown to be slightly higher than first-order accurate in space and scales well on hundreds of processors on high-performance parallel computing platforms. To enforce no-slip boundary conditions on particle surfaces, a tuning parameter, the number of outer forcing loops $n_f$, was introduced and calibrated based on experiments. The accuracy of fluid-particle interactions was validated against various test cases corresponding to a single particle settling in a periodic domain with different terminal settling velocities. These cases showed that the tuning parameter $n_f \ge 1$. Since the Immersed Boundary Method does not resolve the fluid-particle interactions when particles are very close to or in contact with one another, collision models are implemented to simulate the interactions between particles and flow. The normal contact model we implement is based on the adaptive collision time step model proposed by Kempe and Fr\"{o}hlich~~\cite{Kempe2012-pl} to remove the constraint resulting from the collision time-step size which is typically much smaller than the fluid time-step size. The tangential contact model utilizes a spring-dash-pot model~\cite{Biegert2017-ku} while the lubrication model is modified based on an algebraic relationship. To ensure accurate collisions between particles, specification of the minimum separation distance between particles, $\zeta_{min}$, is needed. Since $n_f$ affects both fluid-particle and particle-particle interactions, $n_f$ and $\zeta_{min}$ must be calibrated to reproduce experimental results of particle-wall interactions. The collision models are calibrated with experimental results of a single particle colliding with a wall, in which the rebound height of the particle matches the experimental rebound trajectories with tuning parameters $n_f=2$ and $\zeta_{min}=3.0\times 10^{-3}r_p$. Our results also show that a grid resolution of $d_p/h=20$ can produce results that closely match the experiments. These parameters represent a good balance between computational cost and accuracy. Finally, our collocated-grid simulation framework for polydispersed particles was used to simulate both monodispersed and bidispersed fluidized-bed expansion, demonstrating its capability in reproducing the power law relationship between superficial velocity and porosity. 

\section*{Acknowledgments}
This work used the Extreme Science and Engineering Discovery Environment (XSEDE), which is supported by National Science Foundation grant number ACI-1548562. Simulations were conducted with supercomputer resources under XSEDE Project CTS190063. The authors acknowledge the Texas Advanced Computing Center (TACC) at The University of Texas at Austin for providing HPC resources that have contributed to the research results reported within this paper. This work was funded by the California Energy Commission (CEC) under CEC project number EPC-16-017, the U.S. NSF Engineering Center for Reinventing of the Nation’s Urban Water Infrastructure (ReNUWIt) under Award No. 1028968, and Office of Naval Research Grant N00014-16-1-2256. Bernhard Vowinckel gratefully acknowledges the support through the German Research Foundation, (DFG) grant VO2413/2-1. This paper was prepared as a result of work sponsored in part by the California Energy Commission. It does not necessarily represent the views of the Energy Commission, its employees, or the State of California. Neither the Commission, the State of California, nor the Commission’s employees, contractors, or subcontractors makes any warranty, express or implied, or assumes any legal liability for the information in this paper; nor does any party represent that the use of this information will not infringe upon privately owned rights. This paper has not been approved or disapproved by the Commission, nor has the Commission passed upon the accuracy of the information in this paper.

\end{document}